\documentclass[11pt,a4paper]{article}

\usepackage{amsmath,amsthm,amsfonts,amssymb,color,hyperref,anysize,enumitem,graphicx,epstopdf,algorithm,algpseudocode,cleveref,multirow,url}
\usepackage[usenames,dvipsnames]{xcolor}
\usepackage[margin=0.75in]{geometry}

\newcommand{\rr}{\mathbb{R}}
\newcommand{\rrplus}{\rr^+}
\newcommand{\nn}{\mathbb{N}}

\newcommand{\ep}{\epsilon}

\newcommand{\la}{\leftarrow}
\newcommand{\ra}{\rightarrow}

\newcommand{\expect}[1]{\mathbb{E}\left[ #1 \right]}

\renewcommand{\bold}[1]{\textbf{#1}}
\newcommand{\parabold}[1]{\noindent\bold{#1}}

\newcommand{\comm}[1]{~~~~~~\text{(#1)}}
\newcommand{\hide}[1]{}

\newcommand{\poly}{\textsf{poly}}

\DeclareMathOperator*{\argmax}{arg\,max}
\DeclareMathOperator*{\Span}{span}

\newcommand{\difft}[1]{\frac{\mathsf{d} #1}{\mathsf{d} t}}

\usepackage{tcolorbox}
\tcbuselibrary{theorems}

\newtcbtheorem[auto counter]{coloredDEF}{Definition}
{colback=blue!5,colframe=blue!40!black,fonttitle=\bfseries,before={\vspace{0.3cm}},after={\vspace{0.3cm}}}{de}

\newtcbtheorem[auto counter]{coloredNOT}{Notation}
{colback=blue!5,colframe=blue!40!black,fonttitle=\bfseries,before={\vspace{0.3cm}},after={\vspace{0.3cm}}}{no}

\newtcbtheorem[auto counter]{assume}{Assumption}
{colback=green!5,colframe=green!60!black!80,fonttitle=\bfseries,before={\vspace{0.3cm}},after={\vspace{0.3cm}}}{assume}

\newtcbtheorem[auto counter]{coloredLEM}{Lemma}
{colback=red!5,colframe=red!60!black,fonttitle=\bfseries,before={\vspace{0.3cm}},after={\vspace{0.3cm}}}{lm}

\newtcbtheorem[auto counter]{coloredTHM}{Theorem}
{colback=purple!5,colframe=purple!50!black,fonttitle=\bfseries,before={\vspace{0.3cm}},after={\vspace{0.3cm}}}{th}

\newtcbtheorem[auto counter]{coloredPROP}{Proposition}
{colback=purple!5,colframe=purple!50!black,fonttitle=\bfseries,before={\vspace{0.3cm}},after={\vspace{0.3cm}}}{pr}

\newtcbtheorem[auto counter]{coloredCOR}{Corollary}
{colback=purple!5,colframe=purple!50!black,fonttitle=\bfseries,before={\vspace{0.3cm}},after={\vspace{0.3cm}}}{co}

\newtheorem{theorem}{Theorem}
\newtheorem{lemma}[theorem]{Lemma}
\newtheorem{corollary}[theorem]{Corollary}
\newtheorem{proposition}[theorem]{Proposition}
\newtheorem{definition}[theorem]{Definition}

\newcommand{\calL}{\mathcal{L}}

\newcommand{\calO}{\mathcal{O}}

\newcommand{\bbe}{\mathbf{e}}
\newcommand{\bbf}{\mathbf{f}}
\newcommand{\bbg}{\mathbf{g}}
\newcommand{\bbh}{\mathbf{h}}

\newcommand{\bbp}{\mathbf{p}}
\newcommand{\bbq}{\mathbf{q}}
\newcommand{\bbr}{\mathbf{r}}

\newcommand{\bbx}{\mathbf{x}}
\newcommand{\bby}{\mathbf{y}}
\newcommand{\bbz}{\mathbf{z}}

\newcommand{\bbA}{\mathbf{A}}
\newcommand{\bbB}{\mathbf{B}}

\newcommand{\bbI}{\mathbf{I}}

\newcommand{\bbM}{\mathbf{M}}

\newcommand{\bbX}{\mathbf{X}}
\newcommand{\bbY}{\mathbf{Y}}
\newcommand{\bbZ}{\mathbf{Z}}

\newcommand{\inner}[2]{\left\langle ~#1 ~,~ #2~\right\rangle}
\newcommand{\trans}{^{\mathsf{T}}}

\title{Vortices Instead of Equilibria in MinMax Optimization:\\Chaos and Butterfly Effects of Online Learning in Zero-Sum Games}
\author{Yun Kuen Cheung \hspace*{1.5in} Georgios Piliouras\\
Singapore University of Technology and Design
}
\date{}

\newcommand{\expp}[1]{e^{^{#1}}}
\newcommand{\vol}{\mathsf{volume}}
\newcommand{\bbone}{\mathbf{1}}
\newcommand{\barr}{\bar{r}}
\newcommand{\elll}{{_\ell}}

\begin{document}

\maketitle

\begin{quote}
``Strictly competitive games constitute one of the few areas in game theory, and indeed in the social sciences, where a fairly sharp, unique prediction is made. \dots Early experiments failed miserably to confirm the theory \dots  A determined effort to design an experimental test of minimax that \dots (succeeds) \dots was recently made\dots  '' -- Robert Aumann~\cite{eatwell1987new}, 1987.
\end{quote}

\begin{abstract}
We establish that algorithmic experiments in zero-sum games ``fail miserably'' to confirm the unique, sharp prediction of maxmin equilibration.
Contradicting nearly a century of economic thought that treats zero-sum games nearly axiomatically as the exemplar symbol of economic stability,
we prove that no meaningful prediction can be made about the  day-to-day behavior of online learning dynamics in zero-sum games.
Concretely, Multiplicative Weights Updates (MWU)  with constant step-size is \emph{Lyapunov chaotic} in the dual (payoff) space.
Simply put, let's assume that an observer asks the agents playing Matching-Pennies
whether they prefer Heads or Tails (and by how much in terms of aggregate payoff so far).
The range of possible answers consistent with any arbitrary small set of initial conditions blows up exponentially with time
everywhere in the payoff space (Figure~\ref{fig:tornado}).
This result is \emph{robust} both \emph{algorithmically} as well as \emph{game theoretically}:
\begin{itemize}[leftmargin=0.2in]
\item \bold{Algorithmic robustness:} Chaos is robust to agents using any of a general sub-family of Follow-the-Regularized-Leader (FTRL) algorithms,
the well known regret-minimizing dynamics, even when agents mix-and-match dynamics, use different or slowly decreasing step-sizes.
\item \bold{Game theoretic robustness:} Chaos is robust to all affine variants of zero-sum games (strictly competitive games), network variants with arbitrary large number of agents and even to competitive settings beyond these.
\end{itemize}

Our result is in stark contrast with the time-average convergence of online learning to (approximate) Nash equilibrium,
a result widely reported as ``(weak) convergence to equilibrium''.
\end{abstract}

\begin{figure}[htp]
	\centering
	\includegraphics[scale=0.35]{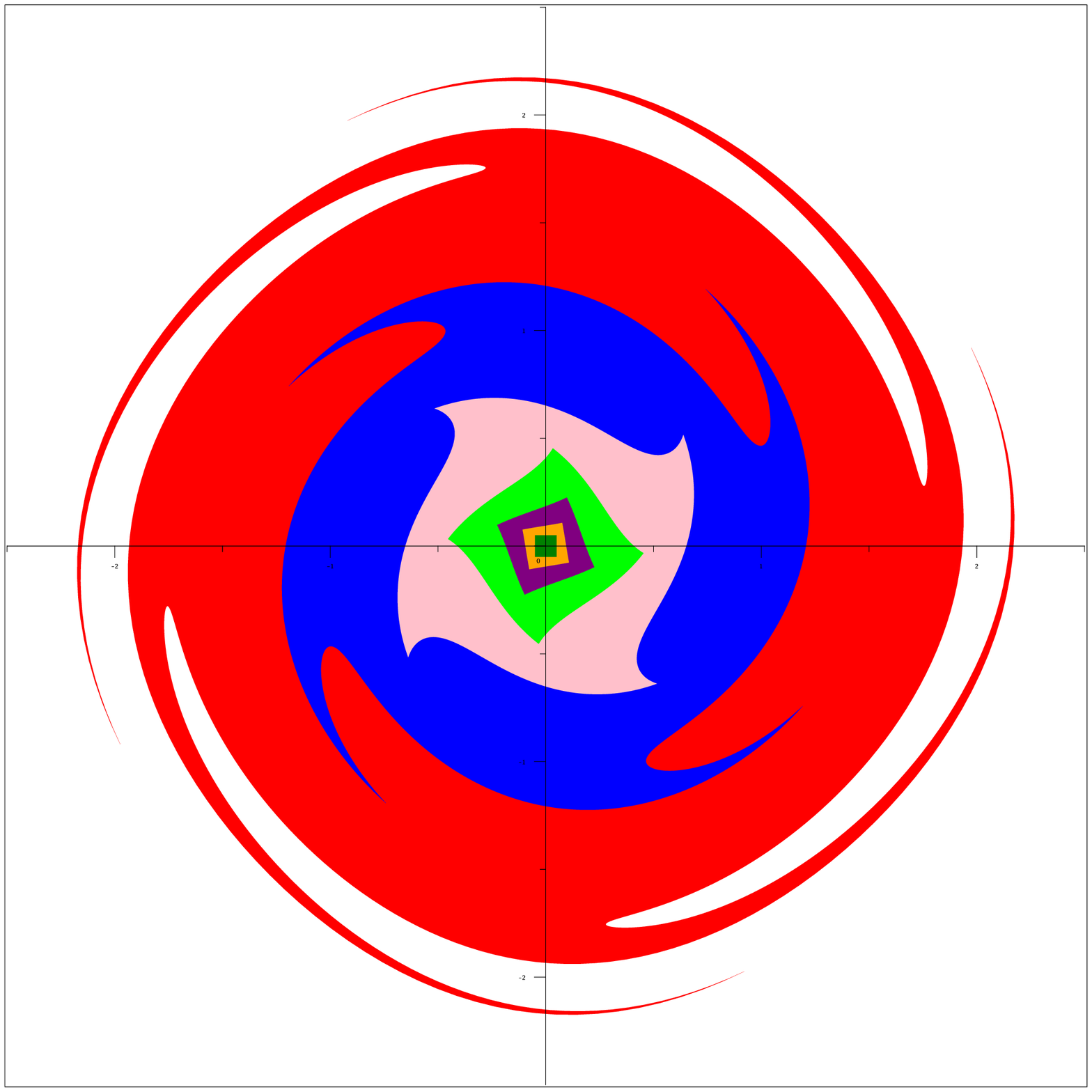}~~~~
	\includegraphics[scale=0.35]{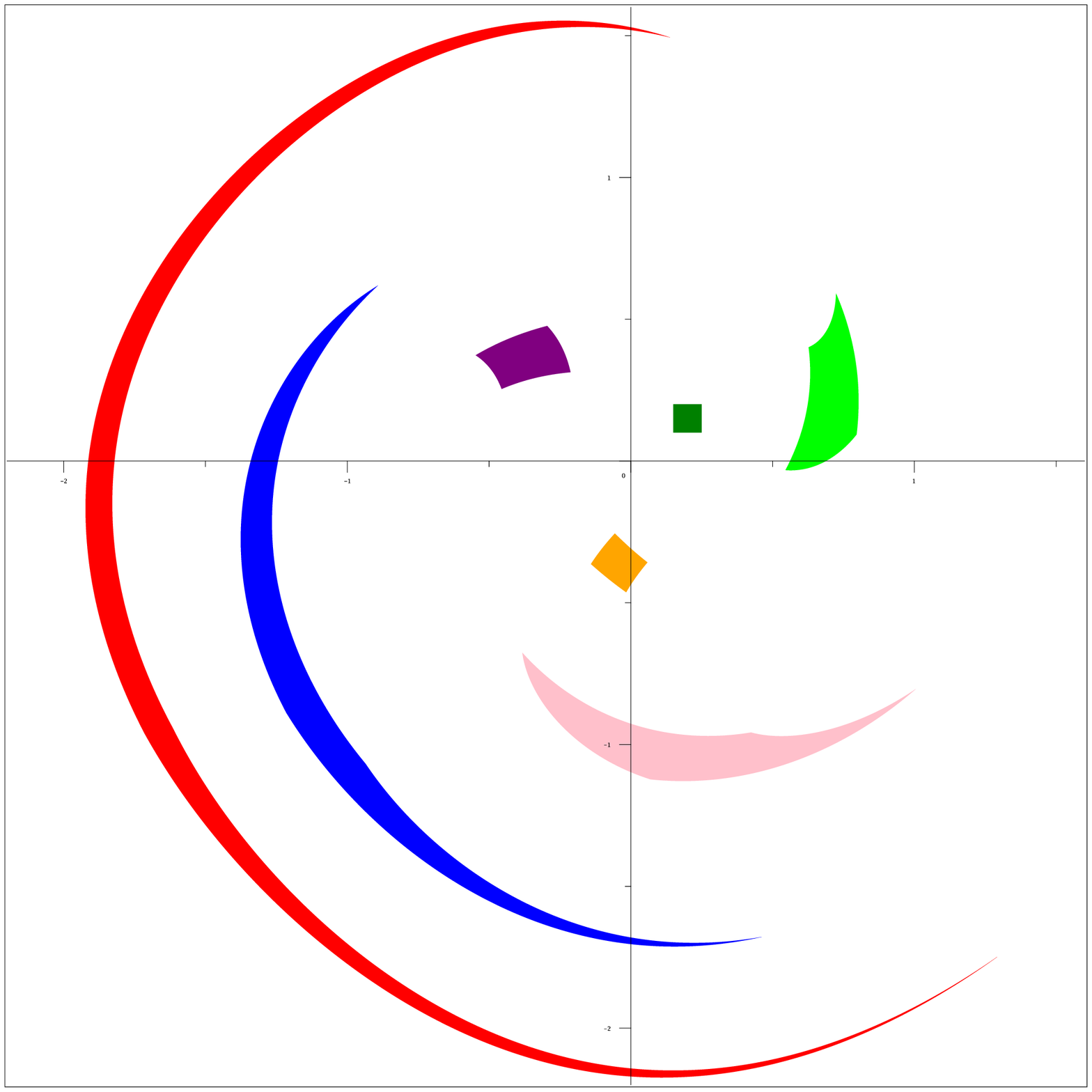}
	\caption{The von Neumann vortex: Volume expansion and chaos of MWU in Matching-Pennies. \hspace{\textwidth} (See Appendix~\ref{app:fig} for explanation.)}
	\label{fig:tornado}
\end{figure}

\parabold{Keywords.}~Lyapunov Chaos, Zero-Sum Games, Multiplicative Weights Updates, Regret Minimization, Gradient Descent Ascent, 
Follow-the-Regularized-Leader Algorithm, Volume of Flow.

\section{Introduction}

Von Neumann's seminal work on zero-sum games \cite{Neumann1928,Neumann1944} set the formal foundations of game theory,
the mathematical theory of coupled strategic behavior. The crowning jewel of his theory is the celebrated minimax theorem
that states that in zero-sum competitions each agent can in isolation compute a safety strategy,
the one that guarantees her, her maxmin payoff and moreover no possible improvement over this minimal guarantee is possible
given that the other agent also plays such a defensive, safety minded strategy.
 
A cornerstone of economic theory, arguably its most resolute thesis,
is that this prescribed solution is indeed the only meaningful behavior in such a setting.
Any rational self-interested learning/adaptive behavior  is bound to gravitate to this benign,
static behavioral snapshot with both agents being deadlocked at their maxmin strategies, or,
minimally even if the system does not equilibrate all of the necessary information needed to understand
the system is represented by these efficiently computable, effectively unique, system states.
  
The number of research threads that follow this kind of reasoning is too numerous to enumerate here,
but they effectively span all disciplines that study the subject,
be it economics, (algorithmic) game theory, online optimization, multi-agent systems, etc.
In fact, the whole sub-field of studying learning dynamics in games started with the work of \cite{Brown1951} and \cite{Robinson1951}
on fictitious play in zero-sum games, which showed that the time-average of the agent behavior converges to their maxmin equilibria.
Ever since that first result a stream of followup works argue convergence of the time-average behavior (strategies/payoffs) of online
(e.g., regret-minimizing) learning dynamics in zero-sum games~\cite{FS1996}.
This line of results represents the main frontier of our understanding  of the effects of rational, self-interested behavior
in strictly competitive settings (see e.g., recent books \cite{young2004strategic,Cesa06,Nisan:2007:AGT:1296179,roughgarden2016twenty}).

As such, the time-average notion has been widely adopted from an algorithmic perspective.
Focusing on time-averages alone, however, can be rather misleading from a behavioral perspective.
For example, in a two-political-party competition,
while the time-average of the political attitude might be moderate which is widely interpreted as good,
at different times it might swing between extremes of the political spectrum which are all often interpreted as bad.
In this context, where such competition can be modelled as zero-sum game,
the theoretical results in~\cite{BP2018,Cheung2018} suggest that online learning
are non-equilibrating, and pushing parties' political attitude towards extremes.
Instability of equilibrium is also a major issue for Generative Adversarial Networks (GANs)~\cite{Goodfellow:2014:GAN:2969033.2969125},
a key application of zero-sum games in AI (see Related Work).
These examples illuminate the importance of developing  a better understanding of the \emph{actual} behavior of game dynamics,
instead of the time-averaged ones. 
As we shall see, our results showcase the possibility of a more unpleasant phenomenon:
online learning in games can be
\emph{chaotic, impossible to predict, at the polar opposite of the picture suggested by the celebrated minimax theorem}.

\smallskip

\parabold{Methodology --- Volume Analysis.}
Our approach is a new methodology in the study of learning in games: we analyze the volume changes of the learning algorithm.
More precisely, given a set of starting points with positive volume (Lebesgue measure),
we analyze the change of the volume as the set is evolved according to the learning algorithm.
In Figure~\ref{fig:tornado}  we plot how a small neighbourhood
around the Nash Equilibrium (NE) (left figure) and one around a non-NE (right figure)
are evolved over time by the MWU algorithm; see Appendix~\ref{app:fig} for more details on how the plots are produced.
%

We show in Sections~\ref{sect:volume-general} and~\ref{sect:two-person-zero-sum} that the volume in the dual (payoff) space increases exponentially,
and as a result in the primal space (probability distributions over strategies) are moving away from Nash and towards the boundary. 
Intuitively, let's assume that an observer asks the agents playing Matching-Pennies whether they prefer Heads or Tails (and by how much in terms of aggregate payoff so far). The range of possible answers consistent with any arbitrary small set of initial conditions blows up exponentially with time everywhere in the payoff space (Figure~\ref{fig:tornado}).
Since the diameter of a set is polynomially lower-bounded by its volume,
our result formally implies \emph{Lyapunov chaos}, a classical notion to measure how chaotic a system is.
It is measured by \emph{Lyapunov time}, which can be informally defined as:
when the starting point is perturbed by a distance of tiny amount of $\delta$,
for how long will the trajectories of the two starting points remain within a distance of at most $2\delta$.
Clearly, the shorter the Lyapunov time, the more chaotic the system is.
We show that the Lyapunov time of MWU in zero-sum game is $\calO(1/\ep^2)$, where $\ep$ is the step-size of the learning algorithm.

This result is \emph{robust} both \emph{algorithmically} as well as \emph{game theoretically}:
\begin{itemize}[leftmargin=0.2in]
\item \bold{Algorithmic robustness:} Chaos is robust to agents using any of a general sub-family of Follow-the-Regularized-Leader (FTRL) algorithms,
the well known regret-minimizing dynamics, even when agents mix-and-match dynamics, use different or slowly decreasing step-sizes (Section \ref{sect:ftrl}). 
\item \bold{Game theoretic robustness:} Chaos is robust to all affine network variants of zero-sum games with arbitrary large number of agents  (Section \ref{sect:graphical-zero-sum}), and even to competitive settings beyond these (Generalized Rock-Paper-Scissors (RPS) games
in Section~\ref{sect:more-games}, and general $2\times 2$ bimatrix game in Section~\ref{app:twobytwo}).
\end{itemize}

\smallskip

\parabold{A Note About Human Behavior in Zero-sum Games.} Our results are in stark contrast with the standard interpretation of the behavior of regret minimizing dynamics in zero-sum games, which is typically referred to as ``converging to equilibrium''. Naturally, we cannot without careful behavioral studies make a claim that human agents in practice adapt their beliefs according to MWU, gradient descent, FTRL, or any other classic first-order optimization method. However, we can confidently deduce a statement in the inverse direction. If as economic theory postulates (and Aumann's quote neatly summarizes) Nash equilibria in zero-sum games are indeed stable and moreover  experimentally verifiable, then this implies that human agents in practice must deviate robustly from the axiomatic perspective of purely optimization driven dynamics as captured by gradient descent and variants and apply carefully tailored equilibrium-seeking behavioral dynamics. Moreover,  this is a cross-cultural behavioral universal.

\smallskip

\smallskip

\parabold{Related Work.} Eshel and Akin~\cite{EA1983} were the first to point out that replicator dynamics are volume-preserving after a transformation. 
In this paper, we will mostly use a slightly different transformation, except for general $2\times 2$ bimatrix games we use their transformation.
Recent work on continuous time dynamics in (variants of) zero-sum games has established that
such dynamics exhibit recurrent, cycle-like behavior, e.g., replicator in network zero-sum games~\cite{PS2014}, 
periodic orbits in team zero-sum games \cite{DBLP:journals/corr/abs-1711-06879} 
and finally recurrence for all FTRL dynamics in affine variants of network zero-sum games~\cite{GeorgiosSODA18}.
Progress in discrete-time dynamics has been much slower but recently based on the above results,
Bailey and Piliouras~\cite{BP2018} and Cheung~\cite{Cheung2018} independently developed non-equilibration analysis
for all FTRL dynamics and MWU dynamics respectively in zero-sum games.

The two prior work~\cite{BP2018,Cheung2018} showed that in a zero-sum game, MWU dynamic diverges from any fully-mixed NE;
more precisely, they showed that the KL-divergence between the current point and the fully-mixed NE strictly increases.
Consequently, the $\omega$-set for any starting point which is not NE must be a subset of the boundary of the strategy space.
The main concern of the prior work is \emph{instability} of the dynamics,
while our current work focuses on \emph{chaos} and \emph{unpredictability}
--- these provide an entirely new, rather intuitive and convincing argument against not only Nash equilibria but
the misconception that zero-sum games are ``easy''.
``Predictability'' is a general target in any branch of science, and particularly so in dynamical systems.
To clarify that ``unpredictability'' is conceptually different from instability, we use the classical example of weather forecast.
It is common sense that weather changes day-to-day, so saying it is unstable is nothing but a tautology.
What is more surprising is its unpredictability (butterfly effect), namely a small change in initial conditions and environmental factors
can lead to significant difference in the outcomes.
Our work is able to spot out and utilise the (geometric) volume measure, which can be viewed as
a summary measure on capturing the effects of perturbation in \emph{all} dimensions.
In contrast, the analyses in~\cite{BP2018,Cheung2018} can be viewed as focusing on an one-dimensional projection (the KL-divergence) of the dynamics,
and clearly had not exploited the richer geometric structures of the dynamics.

By showing that the volumes increase exponentially, a result similar to the $\omega$-set-inside-boundary result in~\cite{BP2018,Cheung2018} can be derived.
One advantage of our approach is it does not need to distinguish between cases on whether a fully-mixed NE exists or not;
however, the statement we can make here will be slightly weaker\footnote{The statement is: in every open subset in the primal space,
there exists a starting point which will eventually get close to the boundary. See Corollaries~\ref{co:dense-two-person-zero-sum} and~\ref{co:dense-two-person-zero-sum-diminish}.}.
A more compelling advantage is that this approach leads to a \emph{global instability} result of NE in RPS games (see Theorem~\ref{th:dense-PRS}).


There have been work reporting observations of chaos even in simple games. Sato et al.~\cite{SAF2002}
focused on a class of RPS games which contains some zero-sum games; the two players employ the continuous-time replicator dynamics.
They ran numerical simulations to find that for those zero-sum games,
the dynamics are chaotic with finite Lyapunov time. Galla and Farmer~\cite{GF2013}
focused on random two-player games where the payoffs to the two players can be positively or negatively correlated;
zero-sum games belong to the negatively correlated regime.
They considered a spectrum of discrete reinforcement learning dynamics, which includes MWU.
Their simulations suggest experimentally that for negatively correlated games MWU exhibit chaos.
We provide a theoretical underpinning for these phenomena for a wide spectrum of dynamics and games.
Palaiopanos et al.~\cite{PPP2017} and Chotibut et al.~\cite{Thip18} studied MWU and its variant in congestion games.
While MWU with very small constant step-size converges to equilibrium in such games,
they showed if we increase the step-size MWU becomes chaotic in a notion first defined by Li and Yorke~\cite{LY1975}. Hence chaotic behavior may be provably verifiable even outside strictly competitive games.

A stream of recent papers proves positive results about convergence to equilibria in (mostly bilinear,  unconstrained) zero-sum games for suitably adapted variants of first-order methods and then apply these techniques to Generative Adversarial Networks (GANs), showing improved performance. One such adapted dynamics are extra-gradient lookahead ``optimistic'' methods \cite{daskalakis2017training}.  
Constrained zero-sum game optimization (e.g. simplex constrained strategies, normal form games) are much harder to address theoretically and only recent work has addressed even the special case of optimistic MWU \cite{daskalakis2018last}.
\cite{2018arXiv180205642B} exploit conservation laws of learning dynamics in zero-sum games (e.g., \cite{PS2014,GeorgiosSODA18}) to develop new algorithms for training GANs that add a new component to the dynamic that aims at minimising this energy function. Different energy shrinking techniques for convergence even in non-convex saddle point problems exploit connections to variational inequalities and employ mirror descent techniques with an extra gradient step \cite{2018arXiv180702629M}. Time-averaging seems to work well in practice for a wide range of architectures, although without necessarily leading to convergence \cite{2018arXiv180604498Y}. Finally, 
\cite{gidel2018negative} provide negative momentum adapted dynamics that add friction to the dynamics. To re-quote Aumann \cite{eatwell1987new}, determined efforts are being made once again to make zero-sum games fit their historically prescribed roles as equilibrium generators, however, zero-sum games are fighting back. For example, optimistic gradient methods as they pressure the system towards stability can end up stabilising even points that are \emph{not} local min-max solutions, i.e., \emph{non-Nash} solutions \cite{daskalakis2018limit}. Our paper showing universal chaos for first order methods in bilinear zero-sum games should be seen as a cautionary tale about the true unpredictability and hardness of training GANs. Not only do we have a long road ahead of us before we have a correct understanding of the behavior of training algorithms for GANs but  more distressingly a thorough understanding might be downright impossible due to emergence of chaos.
\section{Preliminary}\label{sect:prelim}

In this paper, all vectors are denoted by bold lower-case alphabets, and all matrices are denoted by bold upper-case alphabets.
We write $\bbone_n$ for the all-one vector in $\rr^n$, or we simply write $\bbone$ if the dimension is clear from context.
Given a vector $\bbz = (z_1,z_2,\cdots,z_n)$, if $\|\bbz\|_1 = \sum_{j=1}^n z_j \neq 0$,
we say the normalization of $\bbz$ is the vector $\bbz / \|\bbz\|_1$. Let $\Span_{i\in I}(z_i) := \max_{i\in I}(z_i) - \min_{i\in I}(z_i)$.
Let $\Delta^n ~:=~ \left\{ (z_1,z_2,\cdots,z_n) ~\Big|~\forall j\in [n],~z_j\ge 0,~~\text{and}~~\sum_{j=1}^n z_j = 1 \right\}$.
For any positive integer $a$, $[a]$ denotes the set $\{1,2,\cdots,a\}$.

\smallskip

\parabold{Dynamical System, System of Differential Equations and Jacobian.}
A dynamical system is typically described by a system of differential equations over time in $\rr^d$,
governed by $d$ differential equations on the variables $z_1,z_2,\cdots,z_d$,
which are of the form $\difft{z_j} = F_j(z_1,z_2,\cdots,z_d)$, for $j\in [d]$.
Given a starting point $(z_1^\circ,z_2^\circ,\cdots,z_d^\circ)$,
the values of the variables at any time $t\ge 0$ are typically uniquely determined;
precisely, given the starting point, for each $j\in [d]$, there is a function $z_j:\rrplus \ra \rr$
such that altogether they satisfy the system of differential equations, with $(z_1(0),z_2(0),\cdots,z_d(0))$ being the starting point.
The collection of such functions is called the \textit{trajectory} of the given starting point.
The \textit{flow} of a given starting point at time $t$ is simply $(z_1(t),z_2(t),\cdots,z_d(t))$.
In this paper, we assume that the functions $F_j$ are smooth everywhere.

Given a measurable set $S$ and a system of differential equations, the \textit{flow} of $S$ at time $t$
is the collection of the flows of all starting points in $S$ at time $t$;
when the underlying dynamical system is clear from context, we denote it by $S(t)$.
Let $\vol(S)$ denote the Lebesgue volume of a measurable set $S$.
In the rest of this paper, all sets $S$ are assumed to be measurable and bounded.

The \textit{Jacobian} of the system is a $d\times d$-matrix, with the entry in the $i$-th row and $j$-th column be
$\frac{\partial F_i}{\partial z_j}$.

\smallskip

\parabold{Lyapunov Chaos.} In the study of dynamical systems, \textit{Lyapunov chaos} refer generally to following phenomenon in some systems:
a tiny difference in the starting points can yield widely diverging outcomes \emph{quickly}.
A classical measure of chaos is \textit{Lyapunov time}, which can be defined as:
when the starting point is perturbed by a distance of tiny $\delta$,
for how long will the trajectories of the two starting points remain within a distance of at most $2\delta$.

\smallskip

\parabold{Replicator Dynamics.}
In game setting, Replicator Dynamic (RD) is a continuous-time update rule on a probability distribution over strategies.
Such distribution can be naturally denoted by a strategy vector.
Briefly speaking, in RD, the relative change of a probability density in a strategy vector
is same as the payoff from that strategy minus the average payoff at the current probability distribution.

In two-person bimatrix game setting, where the payoffs of the two players are given by matrices $\bbA,\bbB$ respectively,
let the strategy set of Players 1 and 2 be $J$ and $K$ respectively, and let $n = |J|$, $m = |K|$.
We denote a strategy vector of Players 1 and 2 be $\bbx\in \Delta^n$ and $\bby\in \Delta^m$ respectively.

Then RD is governed by the following system of differential equations:
\begin{equation}\label{eq:standard-RD}
\difft{x_j} ~=~ x_j \left[ (\bbA\bby)_j ~-~ \sum_{\ell\in J} x_\elll \cdot (\bbA \bby)_\elll \right]~~~~~~~~~~
\difft{y_k} ~=~ y_k \left[ (\bbB\trans \bbx)_k ~-~ \sum_{\ell\in K} y_\elll \cdot (\bbB\trans \bbx)_\elll \right].
\end{equation}
We follow convention by assuming that every entry in $\bbA,\bbB$ is within the interval $\pm 1$.

\subsection{A Transformation of Replicator Dynamics}

Next, we discuss a crucial transformation of RD system~\eqref{eq:standard-RD}.
This transformation is motivated by the standard implementation of the discrete analogue of RD, the MWU algorithm.

In MWU, for Player 1 and each of her strategies $j$, there is an initial weight $W^\circ_j \in \rr$.
There is no constraint on the initial and subsequent weights.
At time $t$, the weights $(W^t_1,W^t_2,\cdots,W^t_n)$ correspond to a strategy vector $\bbx^t$,
which is the normalization of the vector
\[
\left(~\exp(\ep \cdot W^t_1)~,~\exp(\ep \cdot W^t_2)~,~ \cdots ~,~ \exp(\ep \cdot W^t_n)~\right).
\]
For Player 2, the strategy vector $\bby^t$ is defined similarly.

After each round, the weight of each strategy $j$ is updated by incrementing the value of the payoff to strategy $j$ in that round.
In two-player bimatrix game setting, the update rule is
\begin{equation}\label{eq:MWU-update}
W^{t+1}_j ~~:=~~ W^t_j  ~+~ [\bbA \bby^t]_j.
\end{equation}
Observe that $(W^T_j - W^\circ_j)$ is the cumulative payoff of Player 1
if she were to choose strategy $j$ with certainty in the first $t$ time steps,
while Player 2 were assumed to stick with the choices $\{\bby^t\}_{t=1\cdots T}$.
The weights of Player 2 are updated similarly.

Now we are ready to describe the transformation. The resulting space has dimension $n+m$,
which we call the \textit{dual space} or the \textit{cumulative payoff space}.
The space before transformation will be called the \textit{primal space}.
Let $p_j$ for $j\in [n]$, and let $q_k$ for $k\in [m]$ be the variables in the dual space;
$\bbp$ and $\bbq$ are analogous to the weight vectors of Players 1 and 2 in the MWU algorithm respectively.
We will write $\bbr = (\bbp,\bbq)$.

The transformation from the dual space to the primal space is done via the map
\[
G: ~\bbr = (\bbp,\bbq) ~~\ra~~ \text{concatenation of normalizations of }(\expp{p_1},\expp{p_2},\cdots,\expp{p_n})~~\text{and}~~
(\expp{q_1},\expp{q_2},\cdots,\expp{q_m}).
\]
Observe that $G$ is not one-to-one, but it is easy to see that $G(\bbp_1,\bbq_1) = G(\bbp_2,\bbq_2)$ if and only if
$\bbp_1-\bbp_2 = c_1 \cdot \bbone$ and $\bbq_1-\bbq_2 = c_2 \cdot \bbone$ for some real numbers $c_1,c_2$.

Then consider the system
\begin{equation}\label{eq:second-transform}
\difft{p_j} ~=~ \sum_{\ell\in S_2} A_{j\ell} \cdot \frac{\expp{q_\elll}}{\sum_{z\in S_2} \expp{q_z}}~~~~~~~~~~~~~~~
\difft{q_k} ~=~ \sum_{\ell\in S_1} B_{\ell k}\cdot \frac{\expp{p_\elll}}{\sum_{z\in S_1} \expp{p_z}}.
\end{equation}
We note the similarity between the above system and the MWU update rule~\eqref{eq:MWU-update}.
It is easy to show that system~\eqref{eq:second-transform} is equivalent to the system~\eqref{eq:standard-RD},
as stated precisely in the following proposition.

\begin{proposition}\label{pr:equivalence}
Let $(\bbp_1,\bbq_1)$ and $(\bbp_2,\bbq_2)$ be two vectors in $\rr^{n+m}$ such that
$\bbp_1-\bbp_2 = c_1 \cdot \bbone$ and $\bbq_1-\bbq_2 = c_2 \cdot \bbone$ for some real numbers $c_1,c_2$.
Then the two trajectories obtained from the system~\eqref{eq:second-transform} with starting points $(\bbp_1,\bbq_1)$ and $(\bbp_2,\bbq_2)$
are identical after transformation $G$, which is identical to
the trajectory of the RD system~\eqref{eq:standard-RD} with starting point $G(\bbp_1,\bbq_1) = G(\bbp_2,\bbq_2)$.
\end{proposition}
The system~\eqref{eq:second-transform} is useful since all diagonal entries in its Jacobian are always zero,
a property that leads to volume preservation, which we discuss next.

\subsection{Liouville's Formula and Volume Preservation}\label{subsect:Liouville}

\parabold{Determinant.} Given a $d\times d$ squared-matrix $\bbM$, its determinant is given by the Leibniz formula
$
\det (\bbM) ~=~ \sum_{\sigma \in \mathsf{Perm}([d])} \mathsf{sgn}(\sigma) \cdot \prod_{s=1}^d M_{s,\sigma(s)},
$
where $\mathsf{Perm}([d])$ is the collection of all permutations on $[d]$, and $\mathsf{sgn}(\sigma)$ is the sign of the
permutation $\sigma$. 
Recall from college calculus that determinant computes the signed volume of the parallelepiped spanned by its $d$ column vectors.

The following fact, which follows easily from the Leibniz formula, will be useful. Suppose that the rows and columns of $\bbM$
are indexed by union of two sets $J,K$, and $\bbM$ can be written as
\[
\bbM ~=~ \bbI + \bbZ,~\text{such that}~~~\forall j_1,j_2\in J,~Z_{j_1 j_2} = 0~~~\text{and}~~~\forall k_1,k_2\in K,~Z_{k_1 k_2} = 0.
\]
Furthermore, for any $j\in J,k\in K$, $Z_{jk} = C_{jk} \ep$ for some $C_{jk}\in \rr$,
and  $Z_{kj} = C_{kj} \ep$ for some $C_{kj}\in \rr$.
Then $\det(\bbM)$ is a polynomial of $\ep$ of degree at most $2\cdot \min \{|J|,|K|\}$, and
\begin{equation}\label{eq:det-second}
\det(\bbM) ~~=~~ 1 ~-~ \left(\sum_{j\in J,~k\in K} C_{jk} \cdot C_{kj}\right) \ep^2 ~+~ \calO(\ep^4);
\end{equation}
we note that the coefficient of any odd power of $\ep$ is zero.

\smallskip

\parabold{Liouville's Formula.}
Here, we discuss the necessary ingredient for this paper about Liouville's Formula,
and refer readers to~\cite{Owren2015} for a more elaborate discussion.
Any dynamical system with sum of diagonal entries in its Jacobian always zero is called a \textit{divergence-free} system.

\begin{theorem}\label{th:Liouville}
Let $\dot \bbh = E(\bbh)$ be a system of differential equations on $\rr^d$.
Let $S \equiv S(0)\subset \rr^d$ be a bounded and measurable set.
$\frac{\partial E}{\partial \bbh}$ is the Jacobian of the system.
Then
$
\difft{~\vol(S(t))} = \int_S \mathsf{trace}\left(\frac{\partial E}{\partial \bbh}\right)\,\mathsf{d}V
$.
In particular, if the system is divergence-free, 
then volume is preserved.
\end{theorem}

We will need some elements in a proof of Theorem~\ref{th:Liouville} to proceed.
The proof uses integration for substitution for multi-variables and Taylor expansion.
To apply the former, we need to make sure that for discrete updates, the flow from $S(t)$ to $S(t+1)$ is injective\footnote{This
holds automatically for continuous updates for the flow from $S(t)$ to $S(t+\Delta t)$ for a sufficiently small $\Delta t$,
when $E$ is continuously differentiable and $S(t)$ is bounded.}.
In Appendix~\ref{app:injection}, we prove that $\ep < 1/4$ suffices to guarantee this
for MWU in two-person general-sum game; indeed, the proof covers graphical polymatrix games too, with a smaller upper bound on $\ep$.
The proof uses an appropriate variant of the inverse function theorem.

At time $0$, the solution to the system of ODEs can be locally written as
\begin{equation}\label{eq:ODE-local}
\bbh(t) ~=~ \phi_t(\bbh(0)) ~=~ \bbh(0) ~+~ t \cdot E(\bbh(0)) ~+~ \calO(t^2),
\end{equation}
while volume at time $t$ can be computed by
\[
v(t) ~=~ \int_S \det \left( \frac{\partial \phi_t}{\partial \bbh} \right)\,\mathsf{d}V ~=~ 
\int_S \det \left( \bbI + t\cdot \frac{\partial E}{\partial \bbh} + \calO(t^2) \right)\,\mathsf{d}V,
\]
in which $\bbI$ is the identity matrix.
By expanding the determinant in the RHS, we have
\begin{align*}
v(t) - \vol(S) &~=~ \int_S \left( 1 + t \cdot \mathsf{trace} \left(\frac{\partial E}{\partial \bbh}\right) + \calO(t^2) \right)\,\mathsf{d}V ~-~ 
\int_S 1 \,\mathsf{d}V \\
&~=~ \int_S \left( t \cdot \mathsf{trace} \left(\frac{\partial E}{\partial \bbh}\right) + \calO(t^2) \right)\,\mathsf{d}V.
\end{align*}
Dividing both sides by $t$, and taking the appropriate limit as $t\searrow 0$ completes the proof.

\smallskip

\parabold{Exponentially Increasing Volume.}
Here we focus on MWU discrete-time updates with step-size $\ep$.
We may view a MWU update as equivalent to a continuous-time dynamic in the time interval $[0,\ep]$,
with the function value $E$ \emph{unchanged} during the updates within this time interval.
Consequently, the $\calO(t^2)$ term in~\eqref{eq:ODE-local}, which was to account for the changes in $E$, disappears.
By following the above computations, given a measurable set $S$, and let $S'$ be the flow of $S$ after one time step, we have
\begin{equation}\label{eq:volume-discrete}
\vol(S') ~=~ \int_{s\in S} \det \left( \bbI + \ep\cdot \frac{\partial E(s)}{\partial \bbh} \right) \,\mathsf{d}V.
\end{equation}
If one can show that there exists a $\delta > 0$ such that for all $s\in S$, the integrand is at least $1+\delta$,
then we have $\vol(S') \ge (1+\delta) \cdot \vol(S)$.
If this holds in every time step, then the volume increases exponentially at a rate of at least $(1+\delta)^t$.

\smallskip

\parabold{Interpreting Chaos from Volume Increase.}
Here, we discuss how analyzing volume can be related to chaos and Lyapunov time.
In $\rr^d$ for some fixed $d\in\nn$, the volume of a ball with radius $r$ is strictly less than $2^d r^d$.
Thus, if the volume of a set $S$ is $v$, then the following holds:
for any $s\in S$, there is an $s'\in S$ such that $\|s-s'\| > v^{1/d} / 2$.

If in a dynamical system on $\rr^d$, a set $S$ evolves over time with exponentially increasing volume,
and suppose the rate of increase is at least $(1+\delta)^t$.
Then the following holds: for any starting point $s_0 \in S$, let its flow at time $t$ be $s_t$,
and let $\barr(s_0,t)$ denote the $\ell_2$ distance from $s_t$ to the furthest point in the set $S(t)$.
Then $\barr(d_0,t) = \Omega((1+\delta)^{t/d}) = \Omega((1+\delta/d)^t)$, which is again exponential of $t$.
Consequently, the Lyapunov time is at most $\calO(d/\delta)$.
In all of our results, $\delta = \Theta(\ep^2)$, so the Lyapunov time is at most $\calO(1/\ep^2)$
by considering the dimension $d$ as a fixed constant.
\section{Volume Change of Discrete Multiplicative Weights Updates}\label{sect:volume-general}

Next, we consider the discrete analogue of the system~\eqref{eq:second-transform},
which is exactly the MWU algorithm with step-size $\ep$.
Our calculations in this section are for two-person general-sum games, where $\bbA,\bbB$ can be arbitrary.

Following the notation in Theorem~\ref{th:Liouville}, we rewrite the system~\eqref{eq:second-transform}
as $\dot \bbr = E(\bbr)$, where $\bbr = (\bbp,\bbq)$.
MWU algorithm is then equivalent to the vector-form update rule $\phi(\bbr) ~=~ \bbr + \epsilon \cdot E(\bbr)$.
By~\eqref{eq:volume-discrete}, we are interested in the determinant of the following matrix:
\[
\bbM ~\equiv~ \bbM(\bbr) ~:=~ 
\bbI + \ep\cdot \frac{\partial E}{\partial \bbr}.
\]
Observe that $\frac{\partial E}{\partial \bbr}$ has the properties of the matrix $\bbZ$ appeared in Section~\ref{subsect:Liouville}.
Thus, $\det(\bbM)$, which is the integrand in~\eqref{eq:volume-discrete},
is of the form $1+C(\bbr)\cdot \ep^2+\calO(\ep^4)$, where $C(\bbr)$ can be computed using~\eqref{eq:det-second}.
Hence, when $\ep$ is sufficiently small, the value of $C(\bbr)$ will be decisive for volume change.
Clearly, $C(\bbr)$ is a function of $\bbr$, but we shall see that it is actually a function of $G(\bbr)$, the corresponding primal variables.

Next, we compute $C(\bbr)$ explicitly for two-person general-sum games.
Recall that $\bbM$ is a $(J\cup K) \times (J\cup K)$ squared matrix.
Let $(\bbx,\bby) = G(\bbr)$. 
Due to the structure of $\frac{\partial E}{\partial \bbr}$,
all diagonal entries of $\bbM$ are $1$.
For any distinct $j_1,j_2\in J$ and distinct $k_1,k_2\in K$, $M_{j_1 j_2},M_{k_1,k_2} = 0$.
For $j\in J$ and $k\in K$, we have
\begin{equation}
M_{jk} = \ep \cdot \frac{\partial}{\partial q_k} \left(\frac{\sum_{\ell\in K} A_{j\ell} \cdot\expp{q_\ell}}{\sum_{z\in K} \expp{q_z}}\right)
= \ep \left(\frac{A_{jk} \cdot \expp{q_k}}{\sum_{z\in K} \expp{q_z}} - \frac{\sum_{\ell\in K} A_{j\ell} \cdot\expp{q_\ell}\cdot \expp{q_k}}{\left(\sum_{z\in K} \expp{q_z}\right)^2}\right)
=  \ep y_k \cdot (A_{jk} - [\bbA \bby]_j).\label{eq:Jacobian-explicit}
\end{equation}
Analogously, $M_{kj} ~=~ \ep x_j \cdot (B_{jk} - [\bbB\trans \bbx]_k)$.
By~\eqref{eq:det-second}, 
\[
C(\bbr) = -\sum_{j\in J,~k\in K} x_j y_k \cdot (A_{jk} - [\bbA \bby]_j) \cdot (B_{jk} - [\bbB\trans \bbx]_k).
\]
As promised, $C(\bbr)$ eventually depends on $(\bbx,\bby) = G(\bbr)$ only, but not explicitly on $\bbr = (\bbp,\bbq)$.
Expanding the RHS yields:
\begin{align}
C(\bbr) ~=~ &- \sum_{j\in J,~k\in K} x_j y_k A_{jk} B_{jk}  ~+~ \sum_{j\in J} x_j \cdot [\bbA \bby]_j \cdot \sum_{k\in K} y_k B_{jk}\nonumber\\
&~~~~~~~~~~~+~ \sum_{k\in K} y_k \cdot [\bbB\trans \bbx]_k \cdot \sum_{j\in J} x_j A_{jk}
~-~ \left(\sum_{j\in J} x_j \cdot [\bbA \bby]_j \right) \left( \sum_{k\in K} y_k \cdot [\bbB\trans \bbx]_k \right).\label{eq:second-order}
\end{align}

\section{Exponentially Increasing Volume in Two-Person Zero-sum Games}\label{sect:two-person-zero-sum}

\begin{lemma}\label{lm:non-negative-C}
In any two-person zero-sum game $(\bbA,-\bbA)$, at any point $\bbr$ in which each entry is a finite number, $C(\bbr) \ge 0$.
Furthermore, the equality holds if and only if the game matrix $\bbA$ can be written in the following form for some real numbers
$a_1,a_2,\cdots,a_n,b_1,b_2,\cdots,b_m$:
\begin{equation}\label{eq:uninteresting}
\bbA ~=~ 
\begin{bmatrix}
a_1 - b_1 & a_1 - b_2 & \cdots & a_1 - b_m\\
a_2 - b_1 & a_2 - b_2 & \cdots & a_2 - b_m\\
\vdots & \vdots & \ddots & \vdots\\
a_n - b_1 & a_n - b_2 & \cdots & a_n - b_m
\end{bmatrix}.
\end{equation}
\end{lemma}


Before proving the lemma, we point out that a zero-sum game with matrix~\eqref{eq:uninteresting} 
is ``trivial'', since both players have a dominant strategy:
for Player 1, the dominant strategy is $\argmax_{j\in J} a_j$, while the dominant strategy of Player 2 is $\argmax_{k\in K} b_k$.
In this case, the limit behaviour of MWU is easy to derive:
eventually, each player will play exclusively on her own dominant strategy.

\begin{proof}
For a two-person zero-sum game, $B_{jk} = -A_{jk}$. By~\eqref{eq:second-order},
\[
C(\bbr) ~=~ \sum_{j\in J,~k\in K} x_j y_k (A_{jk})^2 ~-~ \sum_{j\in J} x_j \cdot ([\bbA \bby]_j)^2
~-~ \sum_{k\in K} y_k \cdot ([\bbB\trans \bbx]_k)^2
~-~ \left(\sum_{j\in J} x_j \cdot [\bbA \bby]_j \right) \left( \sum_{k\in K} y_k \cdot [\bbB\trans \bbx]_k \right).
\]
We consider an underlying probability distribution where the tuple $(j,k)$ is chosen with probability $x_j y_k$.
Then we can write
\[
C(\bbr) ~=~ \expect{(A_{jk})^2 - ([\bbA \bby]_j)^2 - ([\bbB\trans \bbx]_k)^2} ~-~ \mathbb{E}\Big[[\bbA \bby]_j\Big] \cdot \expect{[\bbB\trans \bbx]_k}.
\]

Next, note that $\mathbb{E}\Big[[\bbA \bby]_j\Big] = \sum_{j\in J} x_j \cdot [\bbA \bby]_j$ is the expected payoff to Player 1,
while $\expect{[\bbB\trans \bbx]_k}$ is the expected payoff to Player 2.
In a two-person zero-sum  game, the two values are negative to each other. Let $v := \mathbb{E}\Big[[\bbA \bby]_j\Big]$.
Then $C(\bbr) ~=~ \expect{(A_{jk})^2 - ([\bbA \bby]_j)^2 - ([\bbB\trans \bbx]_k)^2} + v^2$.
On the other hand, note that
\[
\expect{A_{jk}} ~=~ \sum_{j\in J} x_j \sum_{k\in K} y_k A_{jk} ~=~ \sum_{j\in J} x_j \cdot [\bbA \bby]_j ~=~ v.
\]
Then by the Cauchy-Schwarz inequality,
\[
v^2 ~=~ \expect{A_{jk} - [\bbA \bby]_j +[\bbB\trans \bbx]_k}^2 ~\le~ \expect{\left(A_{jk} - [\bbA \bby]_j + [\bbB\trans \bbx]_k\right)^2}.
\]
The RHS is expanded as below:
\[
\expect{(A_{jk})^2} ~+~ \expect{([\bbA \bby]_j)^2} ~+~ \expect{([\bbB\trans \bbx]_k)^2} ~-~ 2\cdot \expect{A_{jk} \cdot [\bbA \bby]_j} ~+~ 2\cdot \expect{A_{jk} \cdot [\bbB\trans \bbx]_k}
~-~ 2\cdot \expect{[\bbA \bby]_j \cdot [\bbB\trans \bbx]_k}.
\]
Next, we simplify the last three terms in the RHS.
\begin{align*}
\expect{A_{jk} \cdot [\bbA \bby]_j} &~=~ \sum_{j\in J} x_j \cdot [\bbA \bby]_j\cdot  \sum_{k\in K} y_k A_{jk}
 ~=~ \sum_{j\in J} x_j \cdot ([\bbA \bby]_j)^2 ~=~ \expect{([\bbA \bby]_j)^2};\\
\expect{A_{jk} \cdot [\bbB\trans \bbx]_k} &~=~ \sum_{k\in K} y_k \cdot [\bbB\trans \bbx]_k \cdot  \sum_{j\in J} x_j A_{jk}
 ~=~ -\sum_{k\in K} y_k \cdot ([\bbB\trans \bbx]_k)^2 ~=~ -\expect{([\bbB\trans \bbx]_k)^2}.
\end{align*}
Since the underlying distribution is the product distribution induced by $\bbx$ and $\bby$, we also have
\[
\expect{[\bbA \bby]_j \cdot [\bbB\trans \bbx]_k} ~=~ \expect{[\bbA \bby]_j} \cdot \expect{[\bbB\trans \bbx]_k} ~=~ -v^2.
\]
Combining all above yields
\[
v^2 ~\le~ \expect{(A_{jk})^2} ~-~ \expect{([\bbA \bby]_j)^2} ~-~ \expect{([\bbB\trans \bbx]_k)^2} ~+~ 2v^2 ~=~ C(\bbr) + v^2,
\]
completing the proof of the first part of the lemma. 

To prove the second part of the lemma, first note that since the entries in $\bbr$ are all finite numbers,
$(\bbx,\bby) = G(\bbr)$ are fully-mixed.
Thus, in the application of the Cauchy-Schwarz inequality above,
it is tight if and only if $A_{jk} - [\bbA \bby]_j +[\bbB\trans \bbx]_k$ are identical for all $j\in J$ and $k\in K$.
Next, we prove that the latter condition holds if and only if $\bbA$ has the form of~\eqref{eq:uninteresting}:

\smallskip

\noindent $(\Leftarrow)$ By a direction computation, we have
$A_{jk} - [\bbA \bby]_j +[\bbB\trans \bbx]_k = \sum_{\ell\in K} b_\elll y_\elll - \sum_{\ell\in J} a_\elll x_\elll$,
which is identical for all $j\in J,k\in K$.

\smallskip

\noindent $(\Rightarrow)$ Suppose that at some $\bbr$ we have $A_{jk} - [\bbA \bby]_j +[\bbB\trans \bbx]_k = d$ for all $j\in J,k\in K$.
Then each $A_{jk}$ can be written as $d + [\bbA \bby]_j - [\bbB\trans \bbx]_k$.
We are done by setting $a_j = d + [\bbA \bby]_j$ and $b_k = [\bbB\trans \bbx]_k$ in~\eqref{eq:uninteresting}.
\end{proof}

\subsection{Substantial Exponential Factor of Volume Increment}\label{subsect:exp-factor}

Our next target is to show that the second-order coefficient $C(\bbr)$ is bounded away from zero under suitable conditions.
To begin, we first let
\[
R(\delta) ~:=~ \left\{~\bbr ~|~ \text{every entry in }G(\bbr)~\text{is at least }\delta~\right\}.
\]
Observe that for any $\bbr\in R(\delta)$, $(\bbx,\bby) = G(\bbr)$ are fully-mixed,
and every product $x_j y_k\ge\delta^2$.

Recall that $C(\bbr)$ can be zero only when the underlying zero-sum game is trivial.
Thus, naturally, a lower bound on $C(\bbr)$ will depend on the distance between the game matrix $\bbA$ and the family of those trivial matrices.
Accordingly, we consider the parameter
\[
c(\bbA) ~:=~ \min_{a_1,a_2,\cdots,a_n,b_1,b_2,\cdots,b_m\in \rr}~~\Span_{j\in J,k\in K}~~ \left(A_{jk} - a_j + b_k\right).
\]
When applying the Cauchy-Schwarz inequality above, observe that the gap between the inequality,
which is $\left(\expect{\left(A_{jk} - [\bbA \bby]_j +[\bbB\trans \bbx]_k\right)^2} - \expect{A_{jk} - [\bbA \bby]_j +[\bbB\trans \bbx]_k}^2\right)$,
is indeed the variance of the random variable $(A_{jk} - [\bbA \bby]_j +[\bbB\trans \bbx]_k)$,
and is known to be identical to
\[
\expect{\left(A_{jk} - [\bbA \bby]_j +[\bbB\trans \bbx]_k - \expect{A_{jk} - [\bbA \bby]_j +[\bbB\trans \bbx]_k}\right)^2}.
\]
Thus, by the definition of $c(\bbA)$, in $R(\delta)$, 
the gap is at least $\delta^2 \cdot (c(\bbA)/2)^2 = \delta^2 \cdot c(\bbA)^2 / 4$.

After having a concrete lower bound on $C(\bbr)$, we will still need to bound the higher order terms.
Recall that $\det(\bbM)$ can be written in the form $1+C\ep^2 + \calO(\ep^4)$.
We need a more explicit expansion using the Leibniz formula to bound the higher-order terms.

For each $\min\{n,m\} \ge i\ge 2$, there are at most $\binom{n}{i} \cdot \binom{m}{i}\cdot (i!)^2$ terms
in the summation of the Leibniz formula with factor $\ep^{2i}$.
Each of such terms is a product of $2i$ off-diagonal entries of $\bbM$,
while the absolute value of each such entry of $\bbM$ can be bounded by $2\ep$.
Overall, the sum of all terms with factor $\ep^{2i}$ is bounded by 
$\binom{n}{i} \cdot \binom{m}{i}\cdot (i!)^2 \cdot (2\ep)^{2i} ~\le~ (2\ep\sqrt{nm})^{2i}.$
Thus,
\[
\det(\bbM) - 1
~\ge~ C(\bbr) \cdot \ep^2 - \sum_{i=2}^\infty (2\ep\sqrt{nm})^{2i}
~\ge~ \frac{\delta^2 \cdot c(\bbA)^2}{4}\cdot \ep^2 ~-~ \ep^3 \left[ (2\sqrt{nm})^4 \ep + (2\sqrt{nm})^6 \ep^3 + \cdots \right]
\]
When $\ep \le 1/(32n^2m^2)$, we have $(2\sqrt{nm})^4 \ep + (2\sqrt{nm})^6 \ep^3 + \cdots \le 1$.
Consequently, when\\
$\ep \le \min \{1/(32n^2m^2) ~,~ \delta^2 \cdot c(\bbA)^2 / 8 \}$, we have
$\det(\bbM)  ~\ge~ 1+\frac{\delta^2 \cdot c(\bbA)^2}{8}\cdot \ep^2.$

\begin{theorem}\label{th:volume-increase-two-person-zero-sum}
For a fixed $\delta > 0$, let $S \equiv S(0) \subset \rr^{m+n}$ be a measurable set in $R(\delta)$.
Suppose that MWU algorithm is used by both players to play a non-trivial two-person zero-sum game,
with 
\[
\ep \le \min \left\{1/(32n^2m^2) ~,~ \delta^2 c(\bbA)^2 / 8 \right\} ~=: \bar{\ep}(\delta).
\]
Let $\overline{T}$ be a time such that for all $t\in \{0\}\cup [\overline{T}-1]$, $S(t) \subset R(\delta)$.
Then for any $t\in [\overline{T}]$,
\[
\vol(S(t)) ~~\ge~~ \left( 1 + \frac{\delta^2 \cdot c(\bbA)^2}{8}\cdot \ep^2 \right)^t \cdot \vol(S).
\]
Consequently, the Lyapunov time of the system before reaching $R(\delta)$ is at most $\calO(1/(\delta^2 \ep^2))$,
where the hidden constant depends on the game matrix $\bbA$ only.
\end{theorem}

The interpretation of the above theorem is: as long as the MWU algorithm with
some sufficiently small step-size remains in some strict interior of the primal space,
the volume of the flow of MWU in the dual space increases exponentially with a rate of at least $(1+\Theta(\ep^2))^t$.

\subsection{Reaching Boundary: Exponential Lower Bound vs.~Polynomial Upper Bound on Volume}\label{subsect:bdy}

Theorem~\ref{th:volume-increase-two-person-zero-sum} leaves one question: for how long will $S(t)$ stay within $R(\delta)$.
To answer this question, a key observation is that in the dual space,
in every time step of the MWU flow, each entry of $\bbp,\bbq$ will change within the interval $\pm \ep$.
Hence, unconditionally (not only for zero-sum games, but also for general-sum games),
$S(t)$ must be a subset of the following rectangular hyper-box:
\[
\left\{ (\bbp,\bbq) ~|~ \forall j\in J,~\min_{\bbp \in S} p_j^0 - \ep t ~\le~ p_j ~\le~ \max_{\bbp \in S} p_j^0 + \ep t
~~\text{and}~~\forall k\in K,~\min_{\bbq \in S} q_k^0 - \ep t ~\le~ q_k ~\le~ \max_{\bbq \in S} q_k^0 + \ep t \right\}.
\]
Consequently, the volume of $S(t)$ is unconditionally upper bounded by $\calO((\ep t)^{m+n})$;
note that this bound is $\calO(\poly(t))$ by viewing $m,n,\ep$ as fixed parameters.

However, the volume lower bound in Theorem~\ref{th:volume-increase-two-person-zero-sum} is exponential in $t$.
Thus, the upper and lower bounds are incompatible when $t$ gets large,
implying that $S(t)$ can only stay within $R(\delta)$ for at most the first positive root $t$ of the following equation:
\begin{equation}\label{eq:bdy-time}
\left( 1 + \frac{\delta^2 \cdot c(\bbA)^2}{8}\cdot \ep^2 \right)^t \cdot \vol(S) ~=~ \left[ 2\ep t ~+~ \underbrace{\max \left\{ 
\Span_{\bbp\in S,j\in J} p_j^0~,~
\Span_{\bbq\in S,k\in K} q_k^0 \right\}}_{\gamma} \right]^{m+n}.
\end{equation}
Note that when $t=0$, the LHS is strictly less than the RHS, since $S$ is strictly contained in a hypercube of side length $\gamma$.
Also, when $t\nearrow\infty$, the LHS is asymptotically larger than the RHS.
Thus, a positive root of the equation must exist.

\begin{corollary}\label{co:dense-two-person-zero-sum}
For a fixed $\delta$, let $S \equiv S(0) \in \rr^{m+n}$ be a measurable subset in $R(\delta)$.
Suppose that MWU are used by both players to play a zero-sum game which is non-trivial,
with step-size $\ep \le \bar{\ep}(\delta)$.
Then there exists a starting point in $S$ such that before the time of first positive root of~\eqref{eq:bdy-time},
its flow reaches the outside of $R(\delta)$.
Consequently, there is a dense set of starting points in $R(\delta)$ which their flows will eventually reach the outside of $R(\delta)$.
\end{corollary}

Indeed, the same argument holds so long as the volume lower bound on the LHS of~\eqref{eq:bdy-time} is $\omega(t^{m+n})$.
This allows us to generalize to MWU algorithm with diminishing step-sizes. We present the analogous corollary here,
and defer the details to Appendix~\ref{subsect:diminish}.

\begin{corollary}\label{co:dense-two-person-zero-sum-diminish}
For a fixed $\delta$, let $S \equiv S(0) \in \rr^{m+n}$ be a bounded measurable set in the dual space.
Suppose that MWU are used by both players to play a zero-sum game which is not uninteresting,
with diminishing step-size $\{\ep_t\}$ satisfying 
\[
\ep_1 < \frac 14~~~\text{and}~~~\lim_{t\ra\infty} \ep_t < \bar{\ep}(\delta)~~~\text{and}~~~\limsup_{t\ra\infty} \frac{\sum_{\tau=1}^{t} (\ep_\tau)^2}{\log t} > \frac{16(m+n)}{\delta^2 \cdot c(\bbA)^2}.
\]
Then there exists a starting point in $S$ such that its flow will eventually reach the outside of $R(\delta)$;
consequently, there is a dense set of starting points which their flows will eventually reach the outside of $R(\delta)$.
\end{corollary}

The conditions on $\{\ep_t\}$ can be satisfied by a step-size sequence which is asymptotically like those used in regret minimization:
$
\ep_t = \min \left\{ \frac 1{4+\kappa_1} ~,~ \frac{4\sqrt{m+n+\kappa_2}}{\delta \cdot c(\bbA)}\cdot \frac{1}{\sqrt t} \right\} = \Theta\left(\frac 1{\sqrt t}\right),
$
where $\kappa_1,\kappa_2 > 0$.

We note that while Theorem~\ref{th:volume-increase-two-person-zero-sum} is a novel type of result,
Corollaries~\ref{co:dense-two-person-zero-sum} and~\ref{co:dense-two-person-zero-sum-diminish}
are weaker than the main results in~\cite{BP2018,Cheung2018}.
However, the proofs presented here avoids the need to distinguish between games with fully-mixed NE or not.
Corollary~\ref{co:dense-two-person-zero-sum} also provides an explicit time bound (the first positive root of~\eqref{eq:bdy-time})
for reaching the boundary.
More importantly, in Section~\ref{sect:more-games}, we will see that the technique (exponential lower bound vs.~polynomial upper bound)
used for proving the two corollaries can be generalized to prove some novel and interesting results about
some Rock-Paper-Scissors games (for eager readers, please see Theorem~\ref{th:dense-PRS}).
\section{Generalization to the Follow-The-Regularized-Leader Algorithm}\label{sect:ftrl}

\newcommand{\bbpi}{\bbp_i}
\newcommand{\bbxi}{\bbx_i}
\newcommand{\hsi}{h^*_i}
\newcommand{\by}{\bar{y}}
\newcommand{\bx}{\bar{x}}

Recall that in the dual space, the vectors $\bbp,\bbq$ can be viewed as the cumulative payoffs in a two-person game.
Here, we use $\bbpi\in \rr^{n_i}$ to denote the cumulative payoff vector of Player $i$.
The \emph{Follow-The-Regularized-Leader} (FTRL) dynamic with step-size $\ep$
is determined by a convex \textit{regularizer function} $h_i: \Delta^{n_i} \ra \rr$.
In each round, the payoff vector $\bbpi$ is converted to probability distribution by:
\[
\bbxi^t ~\la~ \argmax_{\bbxi\in \Delta^{n_i}} \left\{ ~\inner{\bbpi^t}{\bbxi} ~-~ \frac 1\ep \cdot h_i(\bbxi)~ \right\}.
\]
We note that MWU is a special case of FTRL, by setting $h_i(\bbxi) = \sum_{\ell\in S_i} x_{i\ell} \cdot \ln x_{i\ell}$.

Using this more general update rule, the results presented in this section can extend to settings where
\begin{itemize}[leftmargin=0.2in]
\item players use MWU with different step-sizes: we just need to scale up or down the players' regularizer functions by constant factors;
\item different players using different diminishing-step-sizes,
since our volume analysis can actually permit the regularizer functions be changed over time
(so long as they do not violate the requirements we will impose soon);
recall that Bailey and Piliouras~\cite{BP2018} imposed a requirement on the step-size sequences used by different players;
\item different players mix-and-match dynamics, i.e., the players can use entirely different types of regularizers.
\end{itemize}

In zero-sum games, we want to reproduce an analysis for general FTRL as in Section~\ref{sect:two-person-zero-sum}. 
In full generality it is rather clumsy. Thus, we focus on the special cases when
\begin{itemize}[leftmargin=0.2in]
\item $h_i(\bbxi) = \sum_{j\in S_i} h_{ij}(x_{ij})$ is separable and
second-continuously-differentiable in the relative interior of the primal space;
\item for all $x_{ij} > 0$, $h_{ij}''(x_{ij})$ is strictly positive, i.e., $h_{ij}$ is strictly convex; and
\item the corresponding FTRL dynamic guarantees $\bbxi$ stays
full-mixed.\footnote{This condition holds if $\lim_{x_{ij}\searrow 0} h'_{ij}(x_{ij}) = -\infty$ for all $i,j$.}
\end{itemize}

In Appendix~\ref{app:ftrl}, we show that by letting $\bx_{i\ell} = 1/h''(x_{i\ell})$
and $H := \sum_{\ell\in S_i} \bx_{i\ell}$, we have
\begin{equation}\label{eq:partd-ftrl}
\frac{\partial x_{ij}}{\partial p_{ij}} ~=~ \bx_{ij} - \frac{[\bx_{ij}]^2}{H}
~~~~~~\text{and}~~~~~~
\forall \ell\in S_i\setminus \{j\},~
\frac{\partial x_{i\ell}}{\partial p_{ij}} ~=~ -\frac{\bx_{ij} \bx_{i\ell}}{H}.
\end{equation}

Next, we explain how the previous analyses in this paper can be generalized to FTRL algorithm in two-person zero-sum games.

\medskip

\parabold{Two-person Zero-sum Game.}
Again, 
we focus on showing $C(\bbr)\ge 0$, and defer all other details and result statements to Appendix~\ref{app:ftrl}.
Recall the system~\eqref{eq:second-transform}, which we rewrite here:
\[
\difft{p_j} = \sum_{\ell\in K} A_{j\ell} \cdot y_\elll ~~~\text{and}~~~\difft{q_k} = \sum_{\ell\in J} B_{\ell k}\cdot x_\elll.
\]
Keep in mind that here $\bbx$ is a function of $\bbp$ while $\bby$ is a function of $\bbq$,
which we do not write out explicitly for two reasons:
first, the explicit formula might be complicated, and second, for the need of computing the Jacobian of the system,
knowing \eqref{eq:partd-ftrl} suffices.
As in Section~\ref{sect:volume-general}, in the matrix $\bbM$,
the diagonal entries are all $1$, and $M_{j_1 j_2} = M_{k_1 k_2} = 0$ for any distinct $j_1,j_2\in J$ and distinct $k_1,k_2\in K$.
For $j\in J,k\in K$, by~\eqref{eq:partd-ftrl},
\[
M_{jk} = \ep \cdot \frac{\partial (\sum_{\ell\in K} A_{j\ell} \cdot y_\elll)}{\partial q_k}
= \ep \cdot \left(A_{jk} \by_k - \by_k \cdot \sum_{\ell \in K} A_{j\ell} \cdot \frac{\by_\elll}{\sum_{z\in K} \by_z}\right)
= \ep\by_k \left( A_{jk} - \sum_{\ell\in K} A_{j\ell}\cdot \frac{\by_\elll}{\sum_{z\in K} \by_z} \right).
\]
By comparing the above equality with~\eqref{eq:Jacobian-explicit},
the summation in the RHS of above equality is analogous to $[\bbA \bby]_j$.
%
%
Similarly, $M_{kj} = \ep \bx_j (B_{jk} - \sum_{\ell\in J} B_{\ell k}\cdot \frac{\bx_\elll}{\sum_{z\in J} \bx_z})$,
with the summation here analogous to $[\bbB\trans \bbx]_k$ in the MWU case.

To proceed, we consider 
$
C(\bbr) \left/ \left[\left( \sum_{j\in J} \bx_j \right)\left( \sum_{k\in K} \by_k \right)\right]\right..
$
This quantity is identical to the one given in~\eqref{eq:second-order},
except that each $x_j$ is replaced by $\bx_j / \left( \sum_{j\in J} \bx_j \right)$
and each $y_k$ is replaced by $\by_k / \left( \sum_{k\in K} \by_k \right)$.
Note that coincidentally, the replaced values form two probability distributions over $J$ and $K$ respectively,
which we call them the \emph{shadow distributions} of their corresponding FTRL update rule (see formal definition below).
Therefore, all the arithmetic using expectations that leads to Lemma~\ref{lm:non-negative-C} carries through smoothly
to prove that $C(\bbr) \left/ \left[\left( \sum_{j\in J} \bx_j \right)\left( \sum_{k\in K} \by_k \right)\right]\right.\ge 0$
and hence $C(\bbr)\ge 0$, while equality holds if and only if $\bbA$ has the form~\eqref{eq:uninteresting}.

\begin{definition}\label{de:shadow-distribution}
Given a FTRL update rule with $h_i(\bbxi) = \sum_{j\in S_i} h_{ij}(x_{ij})$ are separable and second-differentiable
in the relative interior of the primal space,
such that the update guarantees $\bbxi$ stays full-mixed. The shadow distribution of the FTRL update rule at $\bbxi$ is
the distribution in which each $j\in S_i$ is realized with probability
\[
\frac{1/h_{ij}''(x_{ij})}{\sum_{\ell\in S_i} 1/h_{i\ell}''(x_{i\ell})}.
\]
\end{definition}

The remaining analyses are spiritually identical to those presented in Sections~\ref{subsect:exp-factor} and~\ref{subsect:bdy},
although some details (e.g., conditions for guaranteeing that a FTRL algorithm is injective) are different.
We defer them to Appendix~\ref{app:ftrl} but present the results here.
Let
\[
\overline{H}(\delta) ~:=~ \max_{\bbr\in R(\delta)}~\max \left\{ \sum_{j\in J} \bx_j~,~\sum_{k\in K} \by_k \right\}.
\]
Also, let $\Delta(\delta)$ denote the minimum possible value in the shadow distributions of any $(\bbx,\bby)=G(\bbr)$,
where $\bbr \in R(\delta)$. We note that $\Delta(\delta)$ is strictly positive for any $\delta$.

\begin{theorem}\label{th:volume-increase-two-person-zero-sum-ftrl}
For a fixed $\delta > 0$, let $S \equiv S(0) \in \rr^{m+n}$ be a measurable subset in $R(\delta)$.
Suppose that MWU algorithm is used by both players to play a non-trivial two-person zero-sum game,
with step-size 
\[
\ep \le \min \left\{~\frac{1}{2\cdot \max\{2,\overline{H}(\delta)\}^4\cdot n^2 m^2} ~,~ \frac{\Delta(\delta)^2 \cdot c(\bbA)^2}{8}
~,~\frac{1}{4\cdot \overline{H}(\delta/2)} ~,~ \frac{\delta}{9} \cdot \min_{\substack{z\ge \delta/2 \\ i\in \{1,2\}\\j\in S_i}} h''_{ij}(z) ~\right\}.
\]
Let $\overline{T}$ be a time such that for all $t\in \{0\}\cup [\overline{T}-1]$, $S(t) \subset R(\delta)$.
Then for any $t\in [\overline{T}]$, 
\[
\vol(S(t)) ~~\ge~~ \left( 1 + \frac{\Delta(\delta)^2 \cdot c(\bbA)^2}{8}\cdot \ep^2 \right)^t \cdot \vol(S).
\]
Consequently, the Lyapunov time of the system before reaching $R(\delta)$ is at most $\calO(1/(\Delta(\delta)^2 \cdot \ep^2))$,
where the hidden constant depends on the game matrix $\bbA$ only.
\end{theorem}

\begin{corollary}\label{co:dense-two-person-zero-sum-ftrl}
For a fixed $\delta$, let $S \equiv S(0) \in \rr^{m+n}$ be a measurable subset in $R(\delta)$.
Suppose that MWU are used by both players to play a zero-sum game which is non-trivial,
with either
\begin{itemize}
\item constant step-size satisfying the bound in Theorem~\ref{th:volume-increase-two-person-zero-sum-ftrl}; or
\item diminishing step-sizes satisfying $\lim_{t\ra\infty} \ep_t = 0$ and
\[
\ep_1 < \min \left\{~\frac{1}{4\cdot \overline{H}(\delta/2)} ~,~ \frac{\delta}{9} \cdot \min_{\substack{z\ge \delta/2 \\ i~\text{and}~j\in S_i}} h''_{ij}(z)~\right\}
~~~\text{and}~~~
\limsup_{t\ra\infty} \frac{\sum_{\tau=1}^{t} (\ep_\tau)^2}{\log t} ~>~ \frac{16(m+n)}{\Delta(\delta)^2 \cdot c(\bbA)^2}
\]
\end{itemize}
Then there exists a starting point in $S$ such that its flow will eventually reach the outside of $R(\delta)$;
consequently, there is a dense set of starting points which their flows will eventually reach the outside of $R(\delta)$.
\end{corollary}
\section{Generalization to Graphical Constant-sum Games}\label{sect:graphical-zero-sum}

\newcommand{\io}{i_1}
\renewcommand{\it}{i_2}
\newcommand{\jo}{j_1}
\newcommand{\jt}{j_2}

\parabold{Graphical Constant-sum Games.}
Next, we consider a striking generalization of two-person zero-sum game, in which there can be many players.
We use $i$ or $i_\bullet$ to denote a player, and $j$ or $j_\bullet$ to denote a strategy.
In a game with $m$ players, we number the players by $1,2,\cdots,m$,
and let $S_i$ denote the strategy set of Player $i$, and $n_i := |S_i|$.
All variables in the primal space are now denoted by $x_{ij}$,
and hence we denote the concatenation of all variables by $\bbx$.
Again, we denote the variables in the dual space by $\bbr$.


A game with $m$ players is a \textit{graphical polymatrix game} if the game is defined as follows: on an undirected graph $H=([m],E_H)$,
each edge $(\io,\it)\in E_H$ corresponds to a bimatrix game between Players $\io$ and $\it$
with strategy sets $S_{\io}$ and $S_{\it}$ respectively.
It is worth noting that the strategy set of a Player $i$ in different bimatrix games is the same,
and every time she plays the game, she must choose the same mixed strategy for all these bimatrix games.
The payoff of a Player $i$ is the sum of payoffs she received from the bimatrix games she involves.
Such a game is a \textit{graphical constant-sum game} if the bimatrix game corresponded by every edge $(\io,\it)\in E$ 
is a two-person constant-sum game (different bimatrix games may have different constants).
WLOG, we assume that each bimatrix game is indeed zero-sum.



\smallskip

\parabold{Analysis.} 
As we have already seen in Section~\ref{sect:two-person-zero-sum}, the key is to show that the value of second-order coefficient $C(\bbr)$
is above zero. The rest of the analysis amounts to bounding $\ep$ to make sure that
the effects of higher order terms are insignificant and $\det(\bbM)$ is strictly positive.

For each zero-sum game corresponding to edge $(\io,\it)$,
let $\bbM_{\io,\it}$ denote the $\bbM$-matrix 
as if there were only these two players playing this zero-sum game.
A crucial observation is:

\begin{center}
\bold{\emph{since the payoffs of a player is simply}}\\
\bold{\emph{the sum of all her payoffs in the bimatrix games which she involves,}}\\
\bold{\emph{if we focus on the sub-squared-matrix of $\bbM$ corresponding to Players $\io$ and $\it$,}}\\
\bold{\emph{it is exactly $\bbM_{\io\it}$.}}
\end{center}

By~\eqref{eq:det-second}, the above observation leads to the following:

\begin{center}
\bold{\emph{the second-order coefficient in $\det(\bbM)$ is}}\\
\bold{\emph{exactly equal to the second-order coefficient in $\sum_{(\io,\it)\in E_H} \det(\bbM_{\io\it})$.}}
\end{center}

Recall that in Lemma~\ref{lm:non-negative-C}, we have already shown that
the second-order coefficient in each $\det(\bbM_{\io\it})$ is non-negative.
Thus, we have proved that $C(\bbr)\ge 0$, while equality holds if and only if
every edge corresponds to a trivial zero-sum game of the form \eqref{eq:uninteresting}.

Accordingly, we say a graphical constant-sum game is non-trivial if at least one of
the two-person constant-sum games corresponded by an edge is non-trivial.

We can define $R(\delta)$ in a similar manner as in Section~\ref{subsect:exp-factor}.
Same as in Section~\ref{subsect:exp-factor}, a strictly positive lower bound on $C(\bbr)$ in $R(\delta)$ can be derived for non-trivial game.
For simplicity, we denote this lower bound by $\bar{C}(\delta)$.

However, unlike in a two-person zero-sum game, the coefficients of $\ep^3$ and other odd powers of $\ep$
in $\det(\bbM)$ can be non-zero. So we need to derive a new bound on higher-order terms.
Let $n = \sum_{i=1}^m n_i$. By expanding the determinant using the Leibniz formula, we have
\[
\det(\bbM) - 1 ~\ge~ \bar{C}(\delta)\cdot \ep^2 - \sum_{i=3}^\infty \binom{n}{i} i! (2\ep)^i
 ~\ge~ \bar{C}(\delta) \cdot \ep^2 - \ep^{2.5} \left[ (2n)^3 \ep^{0.5} + (2n)^4 \ep^{1.5}
+ (2n)^5 \ep^{2.5} + \cdots \right].
\]
When $\ep \le 1/(64n^6)$, we have $(2n)^3 \ep^{0.5} + (2n)^4 \ep^{1.5} + (2n)^5 \ep^{2.5} + \cdots \le \sqrt{2}$.
Consequently, when $\ep \le \min \{1/(64n^6)~,~\bar{C}(\delta)^2 / 8\}$, we have
$
\det(\bbM) ~\ge~ 1 + \frac{\bar{C}(\delta)}{2}\cdot \ep^2.
$

The analysis for diminishing step-sizes can also be extended easily.
Indeed, the only modification needed is to replace the upper bound $1/4$ on $\ep$ to $1/(4\bar{d})$,
where $\bar{d}$ is the maximum degree of the graph underlying the game,
so as to guarantee that MWU is injective and $\bbM$ is strictly diagonally dominant (see Appendices~\ref{app:injection}
and~\ref{subsect:diminish}).

\begin{theorem}\label{th:volume-increase-graphical}
For a fixed $\delta$, let $S \equiv S(0) \in \rr^n$ be a measurable subset in $R(\delta)$.
Suppose that the underlying graphical constant-sum game is non-trivial. Then $\bar{C}(\delta) > 0$.
	
Suppose that MWU are used by all players with step-size satisfying
\[
\ep \le \min \left\{ \frac{1}{64n^6} ~,~\frac{\bar{C}(\delta)^2}{8} \right\}.
\]
Let $\overline{T}$ be a time such that for all $t\in \{0\}\cup [\overline{T}-1]$, $S(t) \subset R(\delta)$.
Then for any $t\in [\overline{T}]$, 
\[
\vol(S(t)) ~~\ge~~ \left( 1 + \frac{\bar{C}(\delta)}{2}\cdot \ep^2 \right)^t \cdot \vol(S).
\]
\end{theorem}

\begin{corollary}\label{co:dense-graphical}
For a fixed $\delta$, let $S \equiv S(0) \subset \rr^n$ be a measurable subset in $R(\delta)$.
Suppose that MWU are used by all players in a non-trivial graphical constant-sum game,
with either
\begin{itemize}
\item constant step-size $\ep \le \min \{1/(64n^6)~,~\bar{C}(\delta)^2 / 8\}$; or
\item diminishing step-size satisfying
\[
\ep_1 < \frac 1{4\bar{d}}~~~~~~\text{and}~~~~~~\lim_{t\ra\infty} \ep_t = 0~~~~~~\text{and}~~~~~~\limsup_{t\ra\infty} \frac{\sum_{\tau=1}^{t} (\ep_\tau)^2}{\log t} ~>~ \frac{4n}{\bar{C}(\delta)},
\]
where $\bar{d}$ is the maximum degree of the graph $H$.
\end{itemize}
Then there exists a starting point in $S$ such that before some finite time,
its flow reaches the outside of $R(\delta)$.
Consequently, there is a dense set of starting points in $R(\delta)$ which their flows will eventually reach the outside of $R(\delta)$.
\end{corollary}
\section{Non-Zero-Sum Games: Generalized Rock-Paper-Scissors Games}\label{sect:more-games}

Consider the Rock-Paper-Scissors (RPS) game with payoff matrices $(\bbA,\bbA\trans)$, where
\begin{equation}\label{eq:PRS}
\bbA ~=~
\begin{bmatrix}
0 & P & -Q\\
-Q & 0 & P\\
P & -Q & 0
\end{bmatrix},\text{ with }P,Q\ge 0.
\end{equation}
This family of games are neither zero-sum nor strictly competitive when $P\neq Q$.\footnote{To see why, assume $P<Q$.
Suppose $\bbx = \bby = (1/3,1/3,1/3)$, then the expected payoffs of both players
are $(P-Q)/3$. However, if both players switch to $(1-2\kappa,\kappa,\kappa)$ for some tiny $\kappa$,
the expected payoffs of both players will be $(4\kappa-2\kappa^2)(P-Q)$,
which is strictly larger than $(P-Q)/3$ when $\kappa$ is sufficiently small.
The case $P>Q$ is symmetric.}

Suppose both players employ MWU to play the game. Since the dimension is small,
computing $\det(\bbM)$ explicitly is easy (say, by using math software).
Let
$C_1 := 2P^2 + 2Q^2 + 5PQ$ and $C_2 := (P-Q)^2$.
Let $a=\expp{p_1}$, $b=\expp{p_2}$, $c=\expp{p_3}$, $d=\expp{q_1}$, $e=\expp{q_2}$ and $f=\expp{q_3}$.
We have
\[
\det(\bbM) ~=~ 1 + \ep^2\cdot 
\left[\frac{C_1 (abdf+abef + acde + acef + bcde + bcdf) - C_2 (abde + acdf + bcef)}{(a+b+c)^2 (d+e+f)^2} \right] ~+~ \calO(\ep^4).
\]
Recall that after transformation $G$, $x_1 = a/(a+b+c)$, $y_1 = d/(d+e+f)$,
and other $x_j$ and $y_k$ can be computed similarly.
Thus, we can rewrite the second-order coefficient $C(\bbr)$ in two different forms:
\begin{align*}
&\left[ C_1 \left(\frac {x_3}{x_2} + \frac {x_3}{x_1}\right) - C_2 \right] x_1x_2y_1y_2 + 
	\left[ C_1 \left(\frac {x_2}{x_1} + \frac {x_2}{x_3}\right) - C_2 \right] x_1x_3y_1y_3 +
	\left[ C_1 \left(\frac {x_1}{x_2} + \frac {x_1}{x_3}\right) - C_2 \right] x_2x_3y_2y_3\\
&\left[ C_1 \left(\frac {y_3}{y_1} + \frac {y_3}{y_2}\right) - C_2 \right] x_1x_2y_1y_2 + 
	\left[ C_1 \left(\frac {y_2}{y_1} + \frac {y_2}{y_3}\right) - C_2 \right] x_1x_3y_1y_3 +
	\left[ C_1 \left(\frac {y_1}{y_2} + \frac {y_1}{y_3}\right) - C_2 \right] x_2x_3y_2y_3.
\end{align*}
A necessary (but not sufficient) condition for $C(\bbr)\le 0$ is both of the followings hold:
\begin{enumerate}
\item[(A)] one of $\frac{x_1}{x_2},\frac{x_2}{x_3},\frac{x_3}{x_1}$ or their reciprocals is less than
$\frac{C_2}{2C_1} = \frac{(1-r)^2}{4 + 10 r + 4r^2}$, where $r:=Q/P$;
\item[(B)] one of $\frac{y_1}{y_2},\frac{y_2}{y_3},\frac{y_3}{y_1}$ or their reciprocals is less than $\frac{C_2}{2C_1}$.
\end{enumerate}

Accordingly, let $W$ denote the collection of all points $(\bbp,\bbq)$ in the dual space
such that the corresponding $(\bbx,\bby)$ satisfy the negations of both (A) and (B).\footnote{Some readers might feel uncomfortable
that we require the negations of both (A) and (B) to hold, since this seems stronger than needed. Our choice is conscious,
as we will need both conditions for giving a good lower bound on $C(\bbr)$.}
Then $x_1\ge 1/(1+4C_1/C_2)$;\footnote{This follows from $x_1 + \frac{2C_1}{C_2}x_1 + \frac{2C_1}{C_2}x_1~\ge~ x_1+x_2+x_3 ~=~ 1$.} the same lower bound holds for other $x_j,y_k$ too.
Thus, in $W$, we can lower bound $C$ by the AM-GM inequality:
\[
\frac{1}{(1+4C_1/C_2)^4} \cdot \left[ C_1 \left( \frac{x_3}{x_2} + \frac{x_3}{x_1} + \frac{x_2}{x_1} + \frac{x_2}{x_3} + \frac{x_1}{x_2} + \frac{x_1}{x_3} \right) - 3C_2 \right]
~\ge~ \frac{6C_1 - 3C_2}{(1+4C_1/C_2)^4}.
\]
Note that this bound is strictly positive as $6C_1 > 3C_2$ always.

By observing that outside $W$, at least one of $x_1,x_2,x_3,y_1,y_2,y_3$ must be strictly less than
$\frac{C_2}{2C_1+C_2}$,\footnote{For instance, say $x_1/x_2 < \frac{C_2}{2C_1}$, then we have $x_1/(1-x_1) < \frac{C_2}{2C_1}$,
which leads to $x_1 < \frac{C_2}{2C_1+C_2}$.}
and by bounding the higher order terms in $\det(\bbM)$ appropriately, we can use the proof technique
behind Corollaries~\ref{co:dense-two-person-zero-sum} and~\ref{co:dense-two-person-zero-sum-diminish} to derive the theorem below.

\begin{theorem}\label{th:dense-PRS}
Suppose two players employ MWU to play RPS game~\eqref{eq:PRS}.
Let $w$ be an interior point in $W$, let $N(w)\subset W$ be a neighbourhood around $w$ with positive volume.
If both players use either
\begin{itemize}
\item constant step-size $\ep$ satisfying
$\ep \le \min\left\{\frac 1{2592}~,~\frac{6C_1 - 3C_2}{2(1+4C_1/C_2)^4}\right\}$; or
\item a sequence of diminishing step-sizes $\{\ep_t\}$ satisfying
\[
\ep_1 < \frac 14~~~\text{and}~~~\lim_{t\ra\infty} \ep_t = 0~~~\text{and}~~~
\limsup_{t\ra\infty} \frac{\sum_{\tau=1}^{t} (\ep_\tau)^2}{\log t} ~>~ \frac{8(1+4C_1/C_2)^4}{2C_1 - C_2},
\]
\end{itemize}
then there exists a finite time $T$ such that the flow of $N(w)$ at time $T$ does not lie entirely within $W$.

Consequently, there is a dense subset of starting points in $W$, such that the flow of each of them
will eventually reach a point such that one of $x_1,x_2,x_3,y_1,y_2,y_3$ is strictly less than $\frac{C_2}{2C_1+C_2}$.
\end{theorem}

We discuss an interpretation of Theorem~\ref{th:dense-PRS}.
Take $w$ be the NE. The table below lists some concrete values of $\frac{C_2}{2C_1+C_2} = \frac{(1-r)^2}{5+8r+5r^2}$
for different values of $r = Q/P$.
Note that all the values are significantly below $\frac 13 \approx 0.333$, the value in all entries of the NE.
Theorem~\ref{th:dense-PRS} implies that the flow of a dense set of starting points in any open neighbourhood of the NE
will eventually get quite far away from the NE.
This is a \emph{global instability} result, in contrast with the classical \emph{local instability} analysis which
linearise the dynamic near the NE locally and compute the unstable and stable manifolds.

\begin{center}
\noindent
\begin{tabular}{ | c | c | c | c | c | c | c | c | c | c | }
\hline			
$r$ & 0.5 & 0.7 & 0.8 & 0.9 & 1.0 & 1.1 & 1.2 & 1.3 & 2\\
\hline
$\frac{C_2}{2C_1+C_2}$ & 0.0244 & 0.00690 & 0.00274 & 0.000615 &  0 & 0.000504 & 0.00183 & 0.00377 & 0.0244\\
\hline  
\end{tabular}
\end{center}

We present a slightly stronger version of the above theorem in Appendix~\ref{app:prs},
which states that for diminishing step-sizes,
the conclusion can actually be improved to: for any $\kappa > 0$, 
two of $x_1,x_2,x_3,y_1,y_2,y_3$ are strictly less than $\frac{C_2}{2C_1+C_2}+\kappa$.
\section{Non-Zero-Sum Games: $\mathbf{2\times 2}$ Bimatrix Games}\label{app:twobytwo}

We consider general $2\times 2$ bimatrix game here.
It is well-known that after a reduction of game matrices, we can consider the following games only:
\[
\bbA ~=~
\begin{bmatrix}
0 & R_1\\
R_2 & 0
\end{bmatrix}~~~~~~
\bbB ~=~
\begin{bmatrix}
0 & R_3\\
R_4 & 0
\end{bmatrix}
\]

In this case, it is more convenient to use a transformation of RD by~\cite{EA1983},
as it will eliminate all $\calO(\ep^3)$ terms.
We describe this transformation for two-person general-sum games.

Number the strategies of Player 1 by $\{1,2\cdots,n\}$ and those of Player 2 by $\{1,2,\cdots,m\}$, where $n,m\ge 2$.
Let $f_j := \ln x_j - \ln x_n$ for $j\in[n-1]$, and let $g_k = \ln y_k - \ln y_m$ for $k\in[m-1]$.
Let $\bbf = (f_1,f_2,\cdots,f_{n-1})$ and $\bbg = (g_1,g_2,\cdots,g_{m-1})$ denote the dual variables.
The dimension of the dual space is $n+m-2$.
We also let $f_n,g_m \equiv 0$, but keep in mind that they are not variables in the dual space.
Note that the variables $\bbx,\bby$ before transformation can be recovered from $\bbf,\bbg$ as follows:
\[
x_j ~=~ \frac{\expp{f_j}}{\sum_{\ell=1}^n \expp{f_\ell}}~~~~~~~~y_k ~=~ \frac{\expp{g_k}}{\sum_{\ell=1}^m \expp{g_\ell}}
\]
Then we have the following form of RD:
\begin{align*}
\difft{f_j} &~=~ \sum_{\ell=1}^m (A_{j\ell} - A_{n\ell}) \cdot \frac{\expp{g_\ell}}{\sum_{z=1}^m \expp{g_z}};\\
\difft{g_k} &~=~ \sum_{\ell=1}^n (B_{\ell k} - B_{\ell m}) \cdot \frac{\expp{f_\ell}}{\sum_{z=1}^n \expp{f_z}}.
\end{align*}
The Jacobian of the system is an $(n+m-2)\times (n+m-2)$-squared matrix with all diagonal entries zero.

Back to $2\times 2$ bimatrix game. The Jacobian is a $2\times 2$ matrix for which we can compute its determinant directly:
\[
\det (\bbM)
~=~ 1  - \expp{f_1} \expp{g_1} (R_1 + R_2) (R_3 + R_4) \ep^2.
\]
In other words, the volume is globally strictly increasing if and only if $(R_1 + R_2) (R_3 + R_4) < 0$.
When $R_1 = -R_2$ or $R_3 = -R_4$, the volume is preserved even with the discrete updates.

One should note that, however, while globally strictly increasing volume implies reaching boundary,
globally strictly decreasing volume does not imply the opposite.
When $R_1 > -R_2 > 0$, and $R_3 + R_4 > 0$, the volume is decreasing,
but since the first strategy of Player 1 is a strictly dominating strategy of her, it follows that $f_1\nearrow \infty$.

The only scenarios when the game has a unique NE which is fully mixed is
when $R_1,R_2$ have the same sign, $R_3,R_4$ have the same sign, and $R_1,R_3$ have different signs.
In this case, volume is globally strictly increasing, indicating the fully mixed NE is globally unstable.
%


%


\section*{Acknowledgements}

Yun Kuen Cheung and Georgios Piliouras acknowledge SUTD grant SRG ESD 2015 097, MOE AcRF Tier 2 Grant 2016-T2-1-170,
grant PIE-SGP-AI-2018-01 and NRF 2018 Fellowship NRF-NRFF2018-07.

\bibliographystyle{plain}
\bibliography{ref,refer}

\newpage

\appendix

\section{Figure~\ref{fig:tornado}}\label{app:fig}

The game used is a classical zero-sum game called Matching-Pennies. The payoff matrix for Player 1 is
\[
\bbA ~=~ \begin{bmatrix}
1 & 0\\
0 & 1
\end{bmatrix}
\]

We use the transformation of~\cite{EA1983}; see Section~\ref{app:twobytwo}.

In the dual space, the Nash equilibrium point is at $(0,0)$, the origin.
We consider two neighbourhoods of starting points. In the left figure,
the neighbourhood is centred at the Nash equilibrium with $\ell_\infty$-radius being $0.05$.
In the right figure, the neighbourhood is centred at the point $(0.2,0.15)$, and again the $\ell_\infty$-radius is $0.05$.

In both cases, we use MWU algorithm with step-size $\ep=0.1$. The evolved sets are coloured dark-green,
orange, purple, lime, pink, blue and red in chronological order.
In the left figure, the evolved sets are captured at times $0,500,1000,1500,2000,2500,3000$ respectively.
In the right figure, the evolved sets are captured at times $0,300,600,900,1200,1500,1800$ respectively.

In the left figure, an outer region strictly contains an inner region, so it shows that the volume expands.
Also, as time goes, the shape of the region goes from square-like to tornado like.

In the right figure, the regions move in clockwise direction around the origin.
Their shapes get thinner, while their diameters grow quickly, indicating that chaos are occurring.
\section{MWU Algorithm is Injective in Graphical Polymatrix Games}\label{app:injection}

All norms we used here are $\ell_\infty$ norms.

Recall that $\bar{d}$ is the maximum degree of the graph underlying the graphical polymatrix game.
For two-person general-sum game, $\bar{d}=1$.

Suppose the contrary that there are two points $\bbr_1$ and $\bbr_2$ such that they map to the same point $\bbr'$ after one round of MWU.
Since the payoff received by each player is within the interval of $\pm \bar{d}$,
we have $\|\bbr_1 - \bbr'\|, \|\bbr_2 - \bbr'\| \le \bar{d}\ep$, and hence
$\|\bbr_1 - \bbr_2\| \le 2\bar{d}\ep$.
Our target is then to derive a contradiction by showing that within the ball
$\overline{B}(\frac{\bbr_1 + \bbr_2}{2},\bar{d}\ep)$, MWU is injective; observe that this ball includes both $\bbr_1,\bbr_2$.
We will use the following version of inverse function theorem~\cite[Theorem 3.1]{Howard1997}:

\begin{theorem}\label{th:IFT}
Let $\bbX,\bbY$ be Banach spaces equipped with norms $\|\cdot \|_{\bbX},\|\cdot \|_{\bbY}$ respectively.
Let $\overline{B}(x_0,r_0)$ be a closed ball in $\bbX$.
Let $f:\overline{B}(x_0,r_0)\ra \bbY$ be a function so that for some invertible linear map $L:\bbX\ra\bbY$ and some $\rho < 1$,
\[
\| L^{-1} \cdot f(x_2) - L^{-1} \cdot f(x_1) - (x_2 - x_1) \|_{\bbX} ~~\le~~ \rho \cdot \|x_2-x_1\|_{\bbX}.
\]
Then $f$ is injective on $\overline{B}(x_0,r_0)$.
\end{theorem}

For our purpose, $\bbX,\bbY$ are identical Euclidean space, both equipped with $\ell_\infty$ norm. Take $L$ to be the identity map.
Then for any $\bbr_a,\bbr_b$ in the ball $\overline{B}(\frac{\bbr_1 + \bbr_2}{2},\bar{d}\ep)$, we have
\begin{align*}
\| L^{-1} \cdot f(\bbr_b) - L^{-1} \cdot f(\bbr_a) - (\bbr_b - \bbr_a) \|
&~=~ \| \bbr_b + \ep \cdot E(\bbr_b) - \bbr_a + \ep \cdot E(\bbr_a) - (\bbr_b - \bbr_a) \|\\
&~=~ \ep \| E(\bbr_b) - E(\bbr_a) \|.
\end{align*}

Suppose that $\|\bbr_a - \bbr_b \| = \kappa$. Then observe that when mapped back to the primal space,
every entry in $\bbx(\bbr_a)$ will be within a multiplicative factor of $e^{2\kappa}$ of the corresponding entry in $\bbx(\bbr_b)$.
Thus, when we focus on the entry of $E$ that corresponds to Player $i$ and her strategy $j$, we have
\begin{align*}
\left| E_{ij}(\bbr_b) - E_{ij}(\bbr_a) \right| &~=~ 
\left|\sum_{(i,\ell)\in E_H} \sum_{k\in S_{\ell}} A^{i\ell}_{jk} \cdot \left[ \bbx_{\ell k}(\bbr_b) - \bbx_{\ell k}(\bbr_a) \right]\right|\\
&~\le~ \sum_{(i,\ell)\in E_H} \sum_{k\in S_{\ell}} 1 \cdot (e^{2\kappa}-1) \bbx_{\ell k}(\bbr_b)\\
&~\le~ \sum_{(i,\ell)\in E_H} (e^{2\kappa}-1)\\
&~\le~ \bar{d} (e^{2\kappa}-1) ~\le~ 4\bar{d} \kappa,
\end{align*}
where the final inequality holds when we assume $\bar{d}\ep < 1/4$, so that $2\kappa \le 4\bar{d}\ep < 1$ and hence $e^{2\kappa}-1\le 4\kappa$.

Consequently, $\ep \| E(\bbr_b) - E(\bbr_a) \| \le 4\bar{d} \ep \kappa$.
For the condition required in the above theorem to hold, it suffices to restrict that $4\bar{d} \ep < 1$, i.e., $\ep < 1/(4\bar{d})$.
\section{Two-person Zero-sum Games}\label{app:zerosum}

\subsection{Diminishing Step-Sizes}\label{subsect:diminish}

Here, we consider the case when the step-sizes used by both players are not constants.
For simplicity, we here assume that both players use the same diminishing step-sizes $\{\ep_t\}$.
Also, we assume that $\ep_1 < 1/4$ and $\lim_{t\ra\infty} \ep_t = 0$.
Let $t_0$ be the first time such that $\ep_t \le \bar{\ep}(\delta)$.
Then the inequality in Theorem~\ref{th:volume-increase-two-person-zero-sum} can be replaced by: for any $t\ge t_0$,
\[
\vol(S(t)) ~~\ge~~ \left[\prod_{\tau=t_0+1}^t \left( 1 + \frac{\delta^2 \cdot c(\bbA)^2}{8}\cdot (\ep_\tau)^2 \right)\right] \cdot \vol(S(t_0)).
\]
To proceed, we need to argue that if $\vol(S(0))$ is strictly positive, then $\vol(S(t_0))$ is also strictly positive.
In Appendix~\ref{app:sdd}, we will prove that when $\ep < 1/4$,
the matrix $\bbM$ is strictly diagonally dominant;
then by a use of Levy-Desplanques theorem, we can show that $\det(\bbM)$ is strictly positive.
Thus, when $S(0)$ has positive measure, $\vol(S(1)) = \int_{S(0)} \det(\bbM)\,\mathsf{d}V$ remains strictly positive.
Inductively, we arrive at the conclusion that $\vol(S(t_0))$ remains strictly positive for any finite $t_0$.

Next, note that
\[
\log \left[\prod_{\tau=t_0}^t \left( 1 + \frac{\delta^2 \cdot c(\bbA)^2}{8}\cdot (\ep_\tau)^2 \right)\right]
~~\ge~~ \frac{\delta^2 \cdot c(\bbA)^2}{16} \cdot \sum_{\tau=t_0}^{t} (\ep_\tau)^2.
\]
If the summation is $\omega(\log t)$, then $\vol(S(t)) = \omega(\poly(t))$.
By the logic identical to that in Section~\ref{subsect:bdy},
the conclusion in Corollary~\ref{co:dense-two-person-zero-sum} applies when the step-sizes are diminishing \emph{gently}.
The next theorem describes the precise conditions required on $\{\ep_t\}$.

%
\subsection{$\bbM$ is Strictly Diagonally Dominant}\label{app:sdd}

For each $j\in J$, $M_{jj} = 1$, for any $j'\in J$ and $j'\neq j$, $M_{jj'} = 0$, and

\[
\sum_{k\in K} M_{jk} ~=~ \sum_{k\in K} \ep y_k (A_{jk} - [\bbA \bby]_j) ~=~
\ep \sum_{k\in K} y_k A_{jk} - \ep [\bbA \bby]_j ~=~ \ep [\bbA \bby]_j - \ep [\bbA \bby]_j = 0.
\]
Thus,
\begin{align*}
\sum_{j'\in J,~j'\neq j} |M_{jj'}| ~+~ \sum_{k\in K} |M_{jk}| &~=~ 2\sum_{k\in K, M_{jk}>0} M_{jk}\\
&~=~ 2\sum_{k\in K, M_{jk}>0} \ep y_k (A_{jk} - [\bbA \bby]_j) ~\le~ 2\sum_{k\in K, M_{jk}>0} \ep\cdot y_k\cdot 2 ~\le~ 4\ep.	
\end{align*}
Consequently, when $\ep < 1/4$, the matrix $\bbM$ is strictly diagonally dominant.

By Levy-Desplanques theorem, $\bbM = \bbM(\ep)$ is non-singular for any $\ep \in [0,1/4)$.
Thus, $\det(\bbM(\ep))$ is non-zero for any $\ep \in [0,1/4)$.

Now, suppose the contrary that $\det(\bbM(\ep')) \le 0$ for some $\ep' \in [0,1/4)$.
Since $\det(\bbM(\ep))$ is a continuous function w.r.t.~$\ep$,
by the intermediate value theorem, there exists an $\ep''\in [0,\ep']$ such that $\det(\bbM(\ep'')) = 0$, a contradiction.
\section{Follow-The-Regularized-Leader Dynamics}\label{app:ftrl}

%

\subsection{Deriving~\eqref{eq:partd-ftrl}}

By standard calculus, we have
\[
\exists~v\in \rr ~\text{such that}~\forall j\in S_i,~p_{ij}^t - h_{ij}'(x_{ij}) ~=~ v.
\]
Suppose we increment $p_{ij}^t$ by a tiny amount $\Delta$, and we want to see how $\bbxi$ changes.
Suppose that $\bbxi$ is changed to $\bbxi^*$, and for each $\ell\in S_i$, $x^*_{i\ell} - x_{i\ell} =: \delta_\ell$.
In the first order arithmetic, we have
\[
\forall \ell \in S_i\setminus \{j\},~~\Delta - h_{ij}''(x_{ij}) \cdot \delta_j = -h_{i\ell}''(x_{i\ell}) \cdot \delta_\ell
~~~~~~\text{and}~~~~~~
\sum_{\ell\in S_i} \delta_\ell = 0.
\]
Note that the above equalities form a linear system with variables $\{\delta_\ell\}$.
It can be solved easily. Let $H := \sum_{\ell\in S_i} 1/h''(x_{i\ell})$. We have
\[
\delta_j ~=~ \frac{(H - 1/h''(x_{ij}))}{H \cdot h''(x_{ij})} \cdot \Delta~~~~~~\text{and}~~~~~~\forall \ell\in S_i\setminus \{j\},~
\delta_\ell ~=~ -\frac{1}{H\cdot h''(x_{ij}) \cdot h''(x_{i\ell}) }\cdot \Delta.
\]
Consequently, we have
\[
\frac{\partial x_{ij}}{\partial p_j} ~=~ \frac{(H - 1/h''(x_{ij}))}{H \cdot h''(x_{ij})}~~~~~~\text{and}~~~~~~
\forall \ell\in S_i\setminus \{j\},~
\frac{\partial x_{i\ell}}{\partial p_j} ~=~ -\frac{1}{H\cdot h''(x_{ij}) \cdot h''(x_{i\ell}) }.
\]

\subsection{FTRL Algorithm is Injective in Two-Person General-Sum Games}

As was done in Appendix~\ref{app:injection}, to show that the algorithm is injective,
it suffices to show that FTRL algorithm is injective inside the ball we introduced in Appendix~\ref{app:injection}.
Here $\bar{d} = 1$. Recall that notation $\bbr = (\bbp,\bbq)$ for two-person general-sum games.

To use Theorem~\ref{th:IFT}, again we take $\bbX,\bbY$ be identical Euclidean space, both equipped with $\ell_\infty$ norm.
This time, $L$ is set to be $u^{-1} \bbI$, for some $u>0$ which we determine later.
Then we have
\[
\| L^{-1}\cdot f(\bbr^b) - L^{-1} \cdot f(\bbr^a) - (\bbr^b - \bbr^a) \| ~\le~ \ep u \| E(\bbr^b) - E(\bbr^a) \| + (1-u) \|\bbr^b - \bbr^a\|.
\]

Suppose that $\|\bbr^b-\bbr^a\| = \kappa$.
To bound the term in RHS, we consider the line segment from $\bbr^a$ to $\bbr^b$, which is parametrized by $[0,1]$.
Recall the matrix $\bbM$ which we computed in Section~\ref{sect:ftrl}. For any $j\in J$, $k\in K$,
$M_{jk}$ is actually $\ep \cdot \frac{\partial E_j}{\partial q_k}$.
Then we have
\begin{align*}
u \left| E_{ij}(\bbr^b) - E_{ij}(\bbr^a) \right| &~=~ u\cdot \int_0^1 \left( \sum_{k\in K} \frac{\partial E_j}{\partial q_k} \cdot (q_{bk} - q_{ak}) \right)\,\mathsf{d}z\\
&~\le~ \kappa u \cdot \int_0^1 \left| \sum_{k\in K} \frac{\partial E_j}{\partial q_k} \right|\,\mathsf{d}z\comm{since $\|\bbr^a-\bbr^b\|_\infty\le \kappa$}\\
&~\le~ \kappa u \cdot \int_0^1 \sum_{k\in K} \left| \by_k \cdot \left(A_{jk} - \sum_{\ell\in K} A_{j\ell}\cdot \frac{\by_\elll}{\sum_{z\in K} \by_z}\right) \right|\,\mathsf{d}z\\
&~\le~ 2\kappa u \cdot \int_0^1 \sum_{k\in K} \by_k\,\mathsf{d}z.
\end{align*}
Thus, by setting $u$ to be the inverse of the maximum value of $\sum_{k\in K} \by_k$ in the ball, we have
\[
u \left| E_{ij}(\bbr^b) - E_{ij}(\bbr^a) \right| \le 2\kappa,~~~\text{and hence}~~~u \| E(\bbr^b) - E(\bbr^a) \| ~\le~ 2\kappa.
\]
Then by restricting $\ep \le u / 4$, we have
\[
\ep u \| E(\bbr^b) - E(\bbr^a) \| + (1-h) \|\bbr^b - \bbr^a\| ~<~ u\kappa/2 + (1-u) \kappa ~=~ (1-u/2)\kappa.
\]
Thus, we can set the parameter $\rho$ in Theorem~\ref{th:IFT} to be $1-u/2$.

\medskip

We still need to provide a concrete value of $u$. Towards this, for any $\delta > 0$, we define
\[
\overline{H}(\delta) ~:=~ \max_{\bbr\in R(\delta)}~\max \left\{ \sum_{j\in J} \bx_j~,~\sum_{k\in K} \by_k \right\}.
\]

Suppose that $\bbr^a,\bbr^b\in R(\delta)$.
It is actually possible that the line segment between $\bbr^a$ and $\bbr^b$ does not fully lie within $R(\delta)$,
i.e., $R(\delta)$ is not convex.
Therefore, $u$ might need to be strictly smaller than $1/\overline{H}(\delta)$.

By~\eqref{eq:partd-ftrl}, to minimize $x_{ij}$, $p_{ij}$ should be pushed to as small as possible, while
$p_{i\ell}$ for $\ell\neq j$ should be pushed to as large as possible.
Therefore, for a fixed $j$, we consider the line segment $\calL$ between the two points $\bbr^a$
and $\bbr^c := \bbr^a - 2\ep \bbe_{ij} + 2\ep \sum_{\ell\in J\setminus \{j\}} \bbe_{i\ell}$,
and parametrized the segment by $[0,1]$.
Since $\bbr^a \in R(\delta)$, $x^a_{ij} \ge \delta$. Then by~\eqref{eq:partd-ftrl},
for any $\bbr^d$ on the line segment, the value of $x^d_{ij}$ can be lower bounded using the following integral for some $\tau\in [0,1]$:
\[
\delta - \int_0^\tau \left(2\ep \bx_{ij} - 2\ep \frac{[\bx_{ij}]^2}{H} + 2\ep \sum_{\ell\in J\setminus \{j\}} \frac{\bx_{ij} \bx_{i\ell}}{H}\right)\,\mathsf{d}z
~~\ge~~ \delta -4\ep \cdot \max_{\bbr\in \calL}~\bx_{ij}(\bbr) ~~=~~ \delta - 4\ep \cdot \max_{\bbr\in \calL}~\frac{1}{h''_{ij}(x_{ij})}.
\]

By imposing that 
\[
\ep \le \frac{\delta}{9} \cdot \min_{z\ge \delta/2} h''_{ij}(z),
\]
the above inequality guarantees that all points in $\calL$ has $x_{ij}$ value at least $\delta /2$.

By assumptions on $h$, the upper bound on $\ep$ is strictly positive, yet it can be arbitrarily close to zero,
since there is nothing to prohibit that $h_{ij}''(z)$ being tiny (but positive) for a particular $z\ge\delta/2$.
For instance, one may
construct a regularizer $h_{ij}$ such that $h_{ij}''(z) \approx 0$ for all $z\ge \alpha$,
but when $z<\alpha$ the value $h_{ij}''(z)$ gets much larger so that $\lim_{z\searrow 0} h'_{ij}(z) = -\infty$.
Then all three conditions on $h$ which we stated in Section~\ref{sect:ftrl} hold.

Of course, the above constructed regularizer is quite unnatural, so one should be able to improve our bounds for a more natural regularizer.
Our key concern here, however, is just to provide a strictly positive upper bound on $\ep$,
for any $\delta > 0$. (The upper bound can depend on $\delta$.)

Anyway, by having the restriction on $\ep$, we can set $u$ to be $\overline{H}(\delta/2)$. In sum, we need the restriction
\[
\ep ~~\le~~ \min \left\{ \frac{1}{4\cdot \overline{H}(\delta/2)} ~,~ \frac{\delta}{9} \cdot \min_{\substack{z\ge \delta/2 \\ i~\text{and}~j\in S_i}} h''_{ij}(z) \right\}.
\]

\subsection{Analysis for Two-person Zero-sum Games}

%

Again, we need to bound the higher order terms.
As in Section~\ref{subsect:exp-factor}, for each $\min\{n,m\}\ge i\ge 2$, there are at most
$\binom{n}{i} \cdot \binom{m}{i}\cdot (i!)^2$ terms in the summation of the Leibniz formula with factor $\ep^{2i}$.
Each of such terms is a product of $2i$ off-diagonal entries of $\bbM$, and its absolute value can be bounded as
\[
\left|\prod_{a=1}^i \prod_{b=1}^i M_{j_a k_b} M_{k_b j_a}\right|
~\le~ \left(\prod_{a=1}^i \bx_{j_a}\right) \cdot \left( \prod_{b=1}^i \by_{k_b} \right)\cdot (2\ep)^{2i}.
\]
Note that all $j_a$'s are distinct, while all $k_b$'s are also distinct.

By the AM-GM inequality, the RHS of the above inequality can be bounded by
\[
\frac{\left( \sum_{a=1}^i \bx_{j_a} \right)^i\left( \sum_{b=1}^i \by_{k_b} \right)^i}{i^{2i}} \cdot (2\ep)^{2i}
~\le~ \left( \overline{H}(\delta) \cdot \ep \right)^{2i}.
\]
Overall, the sum of all terms with factor $\ep^{2i}$ is bounded by 
\[
\binom{n}{i} \cdot \binom{m}{i}\cdot (i!)^2 \cdot (\overline{H}(\delta)\cdot \ep)^{2i} ~\le~ (\overline{H}(\delta)\cdot \ep\sqrt{nm})^{2i}.
\]
Following the calculations in Section~\ref{subsect:exp-factor}, we impose an upper bound of
\[
\ep~~\le~~ \frac{1}{2\cdot \max\{2,\overline{H}(\delta)\}^4\cdot n^2 m^2}.
\]

We will also need the following quantity to bound the gap when applied the Cauchy-Schwarz inequality.
Let $\Delta(\delta)$ denote the minimum possible value in the shadow distributions of any $(\bbx,\bby)=G(\bbr)$,
where $\bbr \in R(\delta)$. We note that $\Delta(\delta)$ is strictly positive for any $\delta$.
Then in $R(\delta)$, we have
\[
C ~\ge~ \frac{\Delta(\delta)^2 c(\bbA)^2}{4}.
\]

Thus, Theorem~\ref{th:volume-increase-two-person-zero-sum} holds for FTRL too, after replacing the upper bound on $\ep$ appropriately.
This yields Theorem~\ref{th:volume-increase-two-person-zero-sum-ftrl}.

\medskip

In Appendix~\ref{subsect:diminish}, we concern MWU with diminishing step-sizes. For FTRL with diminishing step-sizes,
the analysis is essentially the same, except that we need a slightly different argument to show that $\vol(S(t_0))$ is strictly positive.

Consider a matrix $\bbM'$ obtained from $\bbM$ by the following operations:
for each $j\in J$, divide all entries in the $j$-column by $\left( \sum_{\ell\in J} \bx_\ell \right)$, and
for each $k\in K$, divide all entries in the $k$-column by $\left( \sum_{\ell\in K} \by_\ell \right)$.
Note that
\[
\det(\bbM') = \left( \sum_{\ell\in J} \bx_\ell \right)^{-|J|} \left( \sum_{\ell\in K} \by_\ell \right)^{-|K|} \cdot \det(\bbM).
\]
So for showing that $\det(\bbM)$ is strictly positive, it suffices to show that $\det(\bbM')$ is strictly positive.

The advantage of using $\bbM'$ is that it allows us to reuse the calculations in Appendix~\ref{app:sdd}
(by appropriately replacing $\bbx,\bby$ with $\bar{x},\bar{y}$) to show that
when $\ep < 1/(4\cdot \overline{H}(\delta))$,
$\bbM'$ is strictly diagonally dominant and hence $\det(\bbM'),\det(\bbM)$ are both strictly positive.
Thus, Corollary~\ref{co:dense-two-person-zero-sum-diminish} holds for FTRL too,
by replacing the upper bound on $\ep_1$ with appropriately, yielding Corollary~\ref{co:dense-two-person-zero-sum-ftrl}.
\section{A Stronger Theorem for the Generalized Rock-Paper-Scissors Games}\label{app:prs}

We can derive a slightly stronger theorem than Theorem~\ref{th:dense-PRS}.
Here, we only present the result for diminishing step-sizes.

For any $\kappa,\delta > 0$, let $W_{\kappa,\delta}$ denote the collection of all points $(\bbp,\bbq)$ in the dual space
such that the corresponding $(\bbx,\bby)$ satisfy either of the following two conditions:
\begin{itemize}
\item all of $\frac{x_1}{x_2},\frac{x_2}{x_3},\frac{x_3}{x_1}$ and their reciprocals are larger than or equal to
$\frac{C_2}{2C_1} + \kappa$, and at least two of the three entries in $\bby$ are larger than or equal to $\delta$; or
\item all of $\frac{y_1}{y_2},\frac{y_2}{y_3},\frac{y_3}{y_1}$ and their reciprocals are larger than or equal to
$\frac{C_2}{2C_1} + \kappa$, and at least two of the three entries in $\bbx$ are larger than or equal to $\delta$.
\end{itemize}

To understand why $W_{\kappa,\delta}$ is defined as above, we suppose the first condition above holds.
WLOG, assume $y_1,y_2\ge \delta$. Then in the first form of $C$, the second and third terms are non-negative,
while the first term satisfies
\[
\left[ C_1 \left(\frac {x_3}{x_2} + \frac {x_3}{x_1}\right) - C_2 \right] x_1x_2y_1y_2 ~\ge~ 2C_2\kappa \cdot \frac{1}{(1+4C_1/C_2)^2}\cdot \delta^2.
\]
Thus, the RHS of the above inequality can serve as a lower bound for $C$.
Similarly, the same lower bound for $C$ holds if the second condition holds.
Therefore, whenever one of the two conditions hold, we have a strictly positive lower bound for $C$.
Following the logic behind Theorem~\ref{th:dense-PRS}, we have the following theorem.

\begin{theorem}\label{th:dense-PRS-improved}
Suppose two players employ MWU to play the RPS game~\eqref{eq:PRS}.
For any $\kappa,\delta > 0$, let $w$ be an interior point in $W_{\kappa,\delta}$,
let $N(w)\subset W_{\kappa,\delta}$ be a neighbourhood around $w$ with strictly positive volume.
If both players use a sequence of diminishing step-sizes $\{\ep_t\}$ satisfying
\[
\ep_1 < \frac 14~~~~~~\text{and}~~~~~~\lim_{t\ra\infty} \ep_t = 0~~~~~~\text{and}~~~~~~\limsup_{t\ra\infty} \frac{\sum_{\tau=1}^{t} (\ep_\tau)^2}{\log t} ~>~ \frac{12(1+4C_1/C_2)^2}{C_2 \kappa \delta^2},
\]
then there exists a finite time $T$ such that the flow of $N(w)$ at time $T$ does not lie entirely within $W_{\kappa,\delta}$.
Consequently, there is a dense subset of starting points in $W_{\kappa,\delta}$, such that the flow of each of them
will eventually reach a point such that one of the following holds:
\begin{itemize}
\item one of $x_1,x_2,x_3$ is strictly less than $\frac{C_2}{2C_1+C_2} + \kappa$,
and one of $y_1,y_2,y_3$ is strictly less than $\frac{C_2}{2C_1+C_2} + \kappa$; or
\item two of $x_1,x_2,x_3$ are strictly less than $\delta$; or
\item two of $y_1,y_2,y_3$ are strictly less than $\delta$.
\end{itemize}
\end{theorem}

\end{document}